\documentclass[a4paper,11pt]{article}
\pdfoutput=1

\usepackage{jheppub} 
\usepackage{tikz}

\usepackage{subcaption}
\usepackage{simpler-wick}

\preprint{KUNS-3008}

\title{\boldmath Krylov complexity of fermion chain in double-scaled SYK and power spectrum perspective}

\author[a]{Takanori Anegawa,}
\author[b]{Ryota Watanabe}

\affiliation[a]{Yonago College, National Institute of Technology Yonago, Tottori 683-8502, Japan}
\affiliation[b]{Department of Physics, Kyoto University, Kyoto 606-8502, Japan}

\emailAdd{takanegawa@gmail.com}
\emailAdd{watanabe@gauge.scphys.kyoto-u.ac.jp}

\abstract{
We investigate Krylov complexity of the fermion chain operator which consists of multiple Majorana fermions in the double-scaled SYK (DSSYK) model with finite temperature.
Using the fact that Krylov complexity is computable from two-point functions, the analysis is performed in the limit where the two-point function becomes simple and we compare the results with those of other previous studies.
We confirm the exponential growth of Krylov complexity in the very low temperature regime.
In general, Krylov complexity grows at most linearly at very late times in any system with a bounded energy spectrum. Therefore, we have to focus on the initial growth to see differences in the behaviors of systems or operators. Since the DSSYK model is such a bounded system, its chaotic nature can be expected to appear as the initial exponential growth of the Krylov complexity. In particular, the time at which the initial exponential growth of Krylov complexity terminates is independent of the number of degrees of freedom.
Based on the above, we systematically and specifically study the Lanczos coefficients and Krylov complexity using a toy power spectrum and deepen our understanding of those initial behaviors. 
In particular, we confirm that the overall sech-like behavior of the power spectrum shows the initial linear growth of the Lanczos coefficient, even when the energy spectrum is bounded.
}

\begin{document} 
\maketitle
\flushbottom

\section{Introduction}
\label{sec:intro}
The study of chaos and gravity has long played a pivotal role in physics. 
Chaos theory provides a framework for understanding systems susceptible to initial conditions, while gravity governs the large-scale structure of the universe and celestial motion. 
The intersection of these fields is yielding new insights, particularly through concepts like AdS/CFT correspondence and holography \cite{Maldacena:1997re, Gubser:1998bc, Witten:1998qj}.
The relationship between chaos and gravity is prominent in black hole physics. 
Black hole dynamics are thought to be highly chaotic, according to the holographic principle, closely linked to quantum information scrambling.

For quantum systems, operators evolve in time in the Heisenberg picture. 
In complex quantum systems, initially, simple operators are considered to become very complicated over time.
Several quantities have been devised to quantitatively evaluate the complexity of this operator.
The out-of-time-order correlator (OTOC) \cite{1969JETP...28.1200L} quantifies the time evolution of a small perturbation in chaotic quantum systems.
It is considered that the exponential time dependence of the OTOC signals chaoticity.
Its exponent $\lambda_{\rm L}$ is regarded as a quantum counterpart of the classical Lyapunov exponent.
In \cite{Maldacena:2015waa}, it was shown that there exists a universal upper limit, $\lambda_{\rm L}\leq 2\pi T$, for general finite-temperature quantum many-body systems.
This upper bound has the same form as the surface gravity of a black hole, and a quantum system dual to a black hole is expected to saturate this chaos bound \cite{Shenker:2013pqa,Shenker:2013yza,Shenker:2014cwa}.
This chaos bound is also a refinement of the fast scrambling conjecture \cite{Sekino:2008he} that the fastest scrambling time scales as $t_* \sim \log S$, where $S$ is the entropy of the system.
It is also suggested that the thermodynamic well-definedness of the OTOC leads to an upper bound
on the energy dependence of the Lyapunov exponent for general physical system \cite{Hashimoto:2021afd}.

In recent years, Krylov complexity has received much attention as a quantitative measure of operator complexity \cite{Parker:2018yvk}.
It quantitatively evaluates how an operator $\mathcal{O}$ spreads through the operator space, or more precisely, the Krylov subspace.
In the calculation of Krylov complexity, the time evolution of an operator is attributed to the time evolution of a virtual one-dimensional chain system.
The hopping of the chain model is called the Lanczos coefficient, and the Lanczos coefficient governs the time evolution of the operator.
Krylov complexity of an operator can also be completely determined from its auto-correlation function by using the transformation law between the moments of the auto-correlation function and the Lanczos coefficients.
In quantum many-body systems, the Krylov complexity of an operator is expected to grow at most exponentially, and the growth exponent is expected to give an upper bound on the OTOC exponent for that operator.
It has also been proposed to classify phases by using Krylov complexity as an order variable \cite{Anegawa:2024wov}.

In the study of chaos and complexity, The Sachdev-Ye-Kitaev (SYK) model \cite{Sachdev:1992fk, Kitaev2015} has played a central role.
This is a $(0+1)$-dimensional quantum system of $N$ Majorana fermions with Gaussian random interactions.
This model saturates chaos bound at low temperatures and in large $N$ limit, and is considered to be a toy model of a black hole in the context of AdS/CFT correspondence \cite{Kitaev2015, Polchinski:2016xgd, Maldacena:2016hyu}.
In the IR regime, the SYK model is effectively described by the Schwarzian theory and, at the action level, relates to the Jackiw-Teitelboim (JT) gravity model \cite{Jackiw:1984je, Teitelboim:1983ux, Almheiri:2014cka}.
It is also shown that Krylov complexity grows exponentially in time for the single Majorana fermion operator of the SYK model \cite{Parker:2018yvk}.
The growth rate is found to saturate the chaos bound in the low-temperature regime.

Recently, the double-scaled SYK (DSSYK) model has been actively studied since this model allows us to analytically compute several quantities such as the partition function and correlation functions of random fermion chain operators \cite{Erdos:2014zgc, Cotler:2016fpe, Berkooz:2018jqr, Berkooz:2018qkz}.
The DSSYK model is also confirmed to saturate the chaos bound in the regime where the model approaches the conventional low-temperature SYK model.
On the other hand, unlike the conventional SYK model, the DSSYK model is likely to exhibit a property called {\it hyperfast scrambling} in an appropriate parameter region.
This hyperfast growth resembles the volume growth in de Sitter spacetime, and the relationship with de Sitter spacetime is being actively studied \cite{Susskind:2021esx}.

In another direction, the relationship between the DSSYK model and Random Matrix Theories (RMTs) has been studied.
In \cite{Jafferis:2022wez}, the DSSYK model was related to a two-random matrices theory in which the Hamiltonian and the fermion chain are regarded as matrices \cite{Jafferis:2022wez}.
In \cite{Okuyama:2023aup}, the solvable limit of this two-matrix model was discussed.
In particular, the Gaussian Unitary Ensemble appears as the simplest limit.
In this connection, in \cite{Tang:2023ocr}, Krylov complexity for GUE random matrix theory was computed and found to show an initial exponential growth at a rate that saturates the chaos bound.

In \cite{Bhattacharjee:2022ave}, Krylov complexity of a single Majorana fermion operator was studied in the DSSYK model.
However, the behavior of Krylov complexity for the random fermion chain operator has not yet been revealed.
In this paper, we compute Krylov complexity of the fermion chain operators and investigate whether it bounds the time dependence of the OTOC.
We also discuss the hyperfast property in DSSYK from the time dependence of Krylov complexity.
Strictly speaking, in the parameter regime we will deal with, the Lanczos coefficients initially grow linearly and then asymptotically approach a constant value.
In other words, the Krylov complexity initially shows a typical exponential growth.
More importantly, in general, the scrambling time estimated here is independent of the number of degrees of freedom. This is a typical hyperfast behavior.

The behavior of the Krylov complexity or Lanczos coefficients can be characterized by the behavior of the power spectrum, which is a Fourier transform of the auto-correlation function of the operator.
However, conventional discussions of the power spectrum assume a continuous energy spectrum, and a systematic understanding of the power spectrum, including the discrete case, is still insufficient.
In this paper, we also aim to deepen the general and systematic understanding of the behavior of Krylov complexity and Lanczos coefficients by analyzing the toy power spectrum.
This will greatly assist in understanding the behavior of the Krylov complexity and Lanczos coefficient in the DSSYK model.

This paper is organized as follows. In Sec.~\ref{sec:review}, we review the DSSYK model and the relationship between the model and random matrix theories. 
Next, we overview the definition of Krylov complexity. 
In Sec.~\ref{sec:Krylov-in-DSSYK}, we analyze the Krylov complexity of the fermion chain operator of the DSSYK model in various parameter regions. The scrambling time is also discussed.
In Sec.~\ref{sec:toy-model}, we investigate the Lanczos coefficients and Krylov complexity using an toy power spectrum and obtain a systematic understanding of their behavior.
Possible constraints from the physical energy spectrum are also discussed.
Section \ref{sec:discussion} is for the summary and discussion.

\section{Review}
\label{sec:review}
\subsection{Review of the double-scaled SYK model}
\subsubsection{Definition and chord diagrams}
The SYK model is a theory that has received a great deal of attention in the context of low dimensional quantum gravity
It is a model in which $N$ flavors of Majorana fermions $\psi_i\ (i=1,\cdots N)$ have the following interactions
\begin{align}
H=i^{p/2}\sum_{1\leq i_1<\cdots<i_p\leq N}J_{i_1\cdots i_p}\psi_{i_1}\cdots\psi_{i_p}\,.
\end{align}
where $\{\psi_i,\psi_j\}=2\delta_{ij}$.
Here $J_{i_1\cdots i_p}$ is a random coupling constant that follows a Gaussian distribution and satisfies
\begin{align}
\langle J_{i_1\cdots i_p}\rangle = 0\,,\quad \langle J_{i_1\cdots i_p}^2\rangle = \begin{pmatrix}N \\ p\end{pmatrix}^{-1}\mathcal{J}^2\,.
\end{align}
The double-scaled SYK (DSSYK) model is defined with the scaling limit $p \sim \sqrt{N}$. Specifically,
\begin{align}
\label{eq:DS-1}
N \to \infty\,, \quad p \to \infty\,, \quad \lambda \equiv \frac{2 p^2}{N}\ {\rm fixed}\,.
\end{align}
In this setting, \cite{Berkooz:2018jqr} (and related paper \cite{Berkooz:2018qkz}) gave an exact expression for an ensembled partition function using a technique called chord diagram. In the small $\beta$ expansion of the partition function, only even orders survive from Wick's theorem on random averages of coupling constants,
\begin{align}
\langle Z(\beta) \rangle = \langle {\rm Tr}{\,e^{-\beta H}}\rangle = \sum_{k=0}^\infty \frac{(-\beta)^{2k}}{(2k)!}m_k\,.
\label{eq:chord-partition-function}
\end{align}
where $m_k\equiv \langle {\rm Tr}{\,H^{2k}}\rangle $ is called a moment.
By using a chord diagram, the moment $m_{k}$ can be computed as follows. 
First, since Hamiltonian includes Majorana fermions, these traces are approximately the sign of the fermion replacement. When Hamiltonians with different coupling constants swap, the sign swap is $(-1)^k$, where $k$ is the number of fermions the Hamiltonian has in common.
In the double scaling limit, the distribution for this $k$ is the Poisson distribution, and the ensemble average is
\begin{align}
q \equiv \langle (-1)^k \rangle = e^{-\lambda}\,.
\end{align}
Using the above-expected values, the moment is calculated as follows.
\begin{align}
m_k = \mathcal{J}^{2k}\sum_{\pi\in G_{2k}} q^{\chi(\pi)}\,.
\label{eq:analytic-moment}
\end{align}
where $G_{2k}$ is the entire set of chord diagrams with $2k$ points and $\chi(\pi)$ is the number of intersections in the chord diagram $\pi \in G_{2k}$.

\begin{figure}[t]
 \centering
  \begin{tikzpicture}
   \draw (0,0) circle [radius=2];
   \draw[red,very thick] (2,0) arc (-100:-140:5.1);
   \draw[red,very thick] (-2,0) arc (80:40:5.1);
   \draw[red,very thick] ({2*cos(60)},{2*sin(60)}) -- ({2*cos(240)},{2*sin(240)});
   \filldraw (2,0) circle [radius=0.05];
   \filldraw ({2*cos(60)},{2*sin(60)}) circle [radius=0.05];
   \filldraw ({2*cos(120)},{2*sin(120)}) circle [radius=0.05];
   \filldraw (-2,0) circle [radius=0.05];
   \filldraw ({2*cos(240)},{2*sin(240)}) circle [radius=0.05];
   \filldraw ({2*cos(300)},{2*sin(300)}) circle [radius=0.05];
   \end{tikzpicture}
   \caption{Example for $\pi \in G_{6}$ and $\chi(\pi) = 2$.}
\end{figure}
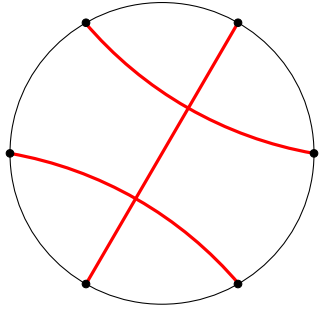

The above is the case of double scaling limit \eqref{eq:DS-1}. The same argument can be made for the more general scaling limit. Specifically, considering the following limit 
\begin{align}
\label{eq:DS-2}
N \to \infty\,, \quad p \to \infty\,, \quad \lambda \equiv \frac{2 p^\alpha}{N}\ {\rm fixed}\,,
\end{align}
the above argument is justified to the extent of $\alpha > \frac{3}{2}$ (See App \ref{app:JPA}).
Using the above method, the moment is computed as follows
\begin{align}
m_k=\int_0^\pi\frac{d\theta}{2\pi}(q,e^{\pm 2i\theta};q)_\infty\left(\frac{2\mathcal{J}\cos\theta}{\sqrt{1-q}}\right)^{2k}\,.
\end{align}
where $(a_1,a_2,\cdots;q)_n$ is $q$-Pochhammer symbol defined by
\begin{align}
(a_1,a_2,\cdots;q)_n =(a_1;q)_n(a_2;q)_n\cdots\,,\quad (a;q)_n=\prod_{k=1}^n(1-aq^{k-1})\,,
\end{align}
and $\pm$ in the equation means multiplying the contributions of all combinations, as in $f(\pm)\equiv f(+)f(-)$. Then, the ensemble-averaged partition function is
\begin{align}
\langle Z(\beta) \rangle
= \int_0^\pi\frac{d\theta}{2\pi}(q,e^{\pm 2i\theta};q)_\infty e^{-\beta E(\theta)}\,,\quad E(\theta)\equiv \frac{2\mathcal{J}\cos\theta}{\sqrt{1-q}}\,,
\label{eq:analytic-partition-function}
\end{align}
where $E(\theta)$ is interpreted as the energy spectrum of the DSSYK model.
From this expression, the density of states as the function of $E$ can be read as
\begin{align}
\rho(E) \equiv \frac{1}{2\pi}\left|\frac{d\theta}{dE}\right|(q,e^{\pm 2i\theta};q)_\infty\,.
\end{align}
Here, in particular, considering the limit of $q\to0\ (\lambda \to \infty)$, only chord diagrams without intersections will contribute to the moment:
\begin{align}
m_k = \mathcal{J}^{2k}\sum_{\pi\in G_{2k}} q^{\chi(\pi)}\to \mathcal{J}^{2k}C_k\,.
\end{align}
where $C_k=\frac{(2k)!}{k!(k+1)!}$ is the Catalan number.
Therefore, in the limit $q\to0$, the partition function can be expressed by using the modified-Bessel function
\begin{align}
\langle Z(\beta) \rangle \to \sum_{k=0}^\infty \frac{(\beta\mathcal{J})^{2k}}{(2k)!}C_k=\frac{I_1(2\beta\mathcal{J})}{\beta\mathcal{J}}\,.
\label{eq:partition-function}
\end{align}
This also can be directly deduced from the \eqref{eq:analytic-partition-function}:
\begin{align}
\langle Z(\beta) \rangle \to \int_0^\pi\frac{d\theta}{2\pi}\,4\sin^2\theta\, e^{-2\beta\mathcal{J}\cos\theta}
= \frac{I_1(2\beta\mathcal{J})}{\beta\mathcal{J}}\,.
\end{align}
Note that, \cite{Berkooz:2018jqr} also points out, the density of states in the limit of $q\to0$ is
\begin{align}
\rho(E) 
&= \frac{1}{2\pi}\times\frac{1}{2\mathcal{J}\sin\theta}\times 4\sin^2\theta = \frac{1}{2\pi\mathcal{J}}\sqrt{4-\frac{E^2}{\mathcal{J}^2}}\,.
\label{eq:Wignerdis}
\end{align}
This is just Wigner semicircle distribution.

On the other hand, the limit of $q\to1$ has also been well studied, sometimes to relate it to the large-$p$ SYK result \cite{Maldacena:2016hyu}.
Then, the moment \eqref{eq:analytic-moment} is just counting the number of elements in $G_{2k}$.\footnote{The high temperature region is implicitly considered here.
The results of the limit $q\to1$ differ slightly depending on whether the temperature is low or high \cite{Berkooz:2018qkz}. Roughly speaking, this may be because at high temperatures, the higher-order chord diagram are suppressed, whereas at low temperatures, the higher-order diagram has some contribution and cannot be ignored.}
Therefore, the partition function \eqref{eq:chord-partition-function} for $q\to1$ becomes\footnote{This is consistent with the large-$p$ SYK result \cite{Maldacena:2016hyu}
\begin{align}
-\beta F/N = \frac{1}{2}\log 2 + \frac{1}{p^2}\pi v\left[\tan\left(\frac{\pi v}{2}\right)-\frac{\pi v}{4}\right]\,,\quad \beta\mathcal{J}_{\rm MS} = \frac{\pi v}{\cos\frac{\pi v}{2}}\,,
\label{eq:MS}
\end{align}
in the high-temperature region. In this region, $\beta\mathcal{J}_{\rm MS}\sim \pi v$ follows and we can obtain
\begin{align}
-\beta F/N \sim \frac{1}{2}\log 2 + \frac{(\beta \mathcal{J}_{\rm MS})^2}{4p^2}\,.
\end{align}
Since the first term on the rhs can be ignored (this is the normalization factor of the $\langle Z(\beta) \rangle$),
\begin{align}
\langle Z(\beta) \rangle \sim \exp{(\beta\mathcal{J}_{\rm MS})^2/2\lambda}\,.
\end{align}
Since $\mathcal{J}$ of \cite{Berkooz:2018jqr,Berkooz:2018qkz} and $\mathcal{J}_{\rm MS}$ of \cite{Maldacena:2016hyu} are related as $\mathcal{J}^2=\mathcal{J}_{\rm MS}^2/\lambda$, the above equation is indeed consistent with \eqref{eq:q-to-1}.}
\begin{align}
\langle Z(\beta) \rangle = \sum_{k=0}^\infty \frac{(\beta\mathcal{J})^{2k}}{(2k)!}\times(2k-1)!! = e^{(\beta\mathcal{J})^2/2}\,.
\label{eq:q-to-1}
\end{align}

\subsubsection{Correlation functions}
Next, let us consider the two-point function of the following {\it fermion chain}
\begin{align}
M\equiv i^{p'/2}\sum_{1\leq i_1<\cdots<i_{p'}\leq N} J'_{i_1\cdots i_{p'}}\psi_{i_1}\cdots\psi_{i_{p'}}\,, \quad
\langle J_{i_1\cdots i_{p'}}'^2\rangle = \begin{pmatrix}N \\ p'\end{pmatrix}^{-1}\,.
\end{align}
where $J'_{i_1\cdots i_{p'}}$ is a random coupling obeying Gaussian distribution and independent of random coupling in the original Hamiltonian.
We can consider for example the two-point function of these operators at finite temperature:
\begin{align}
\langle
{\rm Tr}\,
\wick{
    \c1 M
    (t)
    e^{-\beta H/2}
    \c1 M
    (0)
    e^{-\beta H/2}
}
\rangle
=
\langle
{\rm Tr}\,
\wick{
    \c1 M
    e^{-\beta_1H}
    \c1 M
    e^{-\beta_2H}
}
\rangle\,,
\end{align}
where $M$ denote $M(0)$ and $\beta_1 \equiv \frac{\beta}{2}+it\,,\ \beta_2 \equiv \frac{\beta}{2}-it$.
The contraction symbol with respect to $M$ means the random mean with respect to $J'_{i_1\cdots i_{p'}}$.
According to \cite{Berkooz:2018jqr}, in the double scaling limit with $p' \sim \sqrt{N}$, the two-point function becomes
\begin{align}
\langle
{\rm Tr}\,
\wick{
    \c1 M
    e^{-\beta_1H}
    \c1 M
    e^{-\beta_2H}
}
\rangle
&= \sum_{k_1,k_2}\frac{(-\beta_1)^{k_1}}{k_1!}\frac{(-\beta_2)^{k_2}}{k_2!}
\langle
{\rm Tr}\,
\wick{
    \c1 M
    H^{k_1}
    \c1 M
    H^{k_2}
}
\rangle \notag \\
&= \sum_{k_1,k_2}\frac{(-\beta_1)^{k_1}}{k_1!}\frac{(-\beta_2)^{k_2}}{k_2!}\mathcal{J}^{k_1+k_2}
\sum_{\pi\in G_{k_1,k_2}}q^{\chi_{HH}(\pi)}\tilde{q}^{\chi_{HM}(\pi)}\,,
\label{eq:2pt-chord}
\end{align}
where $\tilde{q}=e^{-2pp'/N}$ is the expectation value of the phase factor that appears when solving for the intersection of the $H$ chord and the $M$ chord.
Here, $G_{k_1,k_2}$ is the whole diagram such that $k_1$ and $k_2$ $H$'s exist on the left and right of the $M$ chord, respectively. Also, $\chi_{HH}(\pi)$ is the number of intersections between $H$ chords in such a diagram $\pi \in G_{k_1,k_2}$ and $\chi_{HM}(\pi)$ denotes the number of intersections between $H$ chord and $M$ chord in $\pi \in G_{k_1,k_2}$.
This is specifically evaluated as follows 
\begin{align}
&\langle
{\rm Tr}\,
\wick{
    \c1 M
    e^{-\beta_1H}
    \c1 M
    e^{-\beta_2H}
}
\rangle \notag \\
&= \int_0^\pi\prod_{j=1}^2\left\{\frac{d\theta_j}{2\pi}(q,e^{\pm 2i\theta_j};q)_\infty\exp\left(-\frac{2\beta_j\mathcal{J}\cos\theta_j}{\sqrt{1-q}}\right)\right\}\frac{(\tilde{q}^2;q)_\infty}{(\tilde{q}e^{i(\pm\theta_1\pm\theta_2)};q)_\infty}\,.
\label{eq:2pt-analytic}
\end{align}

In \cite{Berkooz:2018jqr}, the OTOC of the fermion chain operators is also considered in the low-temperature regime, $\lambda^{3/2}\ll T \ll \lambda^{1/2}$, and small $\lambda\ll 1$.
This is the parameter region of interest for comparison with the large-$p$ SYK model results.
With setting $\mathcal{J}=1$, the Lyapunov exponent is found as $\lambda_{\rm L} = 2\pi T -4\pi \lambda^{-1/2} T^2 + \cdots\,,$ which saturates the chaos bound $\lambda_{\rm L} \leq 2\pi T$ to the leading order.\footnote{Strictly speaking, to compare with the result of the conventional large-$p$ SYK \cite{Maldacena:2016hyu}, it is necessary to note the difference of the normalization of the coupling constant. Since $\mathcal{J}^2=\mathcal{J}_{\rm MS}^2/\lambda$ and $\mathcal{J}=1$, the result of \cite{Berkooz:2018jqr} can be translated as $\lambda_{\rm L}=2\pi T(1-2T/\mathcal{J}_{\rm MS})$ and the temperature regime $T\ll\lambda^{1/2}$ becomes $T\ll \mathcal{J}_{\rm MS}$. This is exactly the same as the result of the large-$p$ SYK model \cite{Maldacena:2016hyu}.}

\subsection{Relationship between RMT and the DSSYK model}
\label{subsec:RMTandDSSYK}
The relationship between the matrix model and the SYK model (and low-dimensional JT gravity theory) has been discussed in various contexts \cite{Cotler:2016fpe}. 

It is known that the relationship between the dynamics of the fermion chain and RMT in the DSSYK model has the following relationship \cite{Jafferis:2022wez}. First, let us consider a two-matrices random matrix model consisting of two matrices $A$ and $B$ as follows
\begin{align}
Z = \int dA dB\ e^{-{\rm Tr} V(A,B)}\,,
\end{align}
where $V(A,B)$ is potential and let the matrix size as $L$. Let $H,M$ denote the Hamiltonian of the DSSYK model and the fermion chain respectively, and correspondence is as follows
\begin{align}
H \leftrightarrow A\,, \quad M \leftrightarrow B\,,
\end{align}
and $L = 2^{N/2}$ with $N$ degrees of freedom in the DSSYK model. Under this correspondence, $V(A,B)$ depends on $q,\tilde{q}$ and $q_M = e^{-2p'^2/N}$ in the DSSYK model.
In \cite{Okuyama:2023aup}, the solvable limit of this two-matrix model is discussed. More ultimately, by setting $q,\tilde{q},q_M\to 0$, this matrix theory becomes just a two-matrixes non-coupled GUE. Therefore, the eigenvalue distribution becomes a Wigner semicircle. This is consistent with \eqref{eq:Wignerdis}.


\subsection{Review of Krylov Complexity}
Here we briefly review the definition of Krylov complexity \cite{Parker:2018yvk}.
For a more extensive review, see \cite{Nandy:2024htc}.
In general, for the operator $\mathcal{O}(t)=e^{iHt}\mathcal{O}_0e^{-iHt}$, its Krylov complexity can be defined by introducing an appropriate inner product in operator space.
It is natural to choose the inner product to be introduced into the operator space according to the form of the two-point function to be considered, and in the case of a two-point function with finite temperature, the following is adopted
\begin{align}
(\mathcal{O}_1|\mathcal{O}_2)\equiv \frac{1}{Z(\beta)}{\rm Tr}\,[e^{-\beta H/2}\mathcal{O}_1^\dagger e^{-\beta H/2}\mathcal{O}_2]\,.
\end{align}
Krylov complexity is specifically defined by the following steps. First, by the Baker-Campbell-Hausdorff formula, we can expand
\begin{align}
\mathcal{O}(t) = \sum_{k=0}^\infty \frac{(it)^k}{k!}\mathcal{L}^k\mathcal{O}_0 \quad (\mathcal{L}\equiv [H,\,\cdot\,])\,,
\end{align}
where we normalize the operator as $(\mathcal{O}_0|\mathcal{O}_0)=1$.
The operator subspace spanned by $\{\mathcal{L}^k\mathcal{O}_0\}$ is called the Krylov subspace.
The Gram-Schmidt orthogonalization (Lanczos method) of $\{\mathcal{L}^k\mathcal{O}_0\}$ appearing here is performed using the inner product of operator spaces.
The algorithm of the orthogonalization is as follows.
\begin{itemize}
\item[1.~] $b_0\equiv0\,, \quad \mathcal{O}_{-1}\equiv0$
\item[2.~] For $n\geq1$: $\mathcal{A}_n=\mathcal{L}\mathcal{O}_{n-1}-b_{n-1}\mathcal{O}_{n-2}$
\item[3.~] Set $b_n=\sqrt{(\mathcal{A}_n|\mathcal{A}_n)}$
\item[4.~] If $b_n=0$ stop; otherwise set $\mathcal{O}_n=\mathcal{A}_n/b_n$ and go to step 2.
\end{itemize}
The $b_n$ are called Lanczos coefficients.
If the dimension $K$ of the Krylov subspace is finite, the above algorithm ends with $b_{K}=0$.
More specifically, if the system under consideration is a $D$-level system, the dimension of the Krylov subspace is known to satisfy $K\leq D^2-D+1$ \cite{Rabinovici:2021qqt}.
This means in particular that the Lanczos coefficient will always be zero in finite steps for finite level systems.
The orthonormal basis $\{\mathcal{O}_n\}$ thus obtained is used to expand $\mathcal{O}(t)$ again:
\begin{align}
\mathcal{O}(t) = \sum_{n=0}^\infty i^n\varphi_n(t)\mathcal{O}_n\,.
\end{align}
The expansion coefficients in this case satisfy the following differential equation
\begin{align}
\partial_t\varphi_n(t) = b_n\varphi_{n-1}(t)-b_{n+1}\varphi_{n+1}(t)
\end{align}
where $b_n$ is Lanczos coefficients.
Once $b_n$ is obtained, the Krylov complexity of $\mathcal{O}(t)$ is defined by
\begin{align}
C_{\rm K}(t) \equiv 1+ \sum_{n=0}^\infty n|\varphi_n(t)|^2
\end{align}
by solving the differential equation above. Here, Krylov complexity is defined so that $C_{\rm K}(0)=1$ for convenience.

Instead of performing the Gram-Schmidt method, there are other methods for obtaining the Lanczos coefficients indirectly from the moments of the two-point function. Consider the auto-correlation function
\begin{align}
C(t) = (\mathcal{O}(t)|\mathcal{O}_0)
\end{align}
of the operator $\mathcal{O}(t)$ and expand it with respect to time:
\begin{align}
C(t) = \sum_{n=0}^\infty \frac{(-1)^n}{(2n)!}\mu_{2n}t^{2n}\,.
\end{align}
The expansion coefficient $\mu_{2n}$ is called the moment. It is known that moments and Lanczos coefficients correspond to each other as in 
\begin{align}
b_1^{2n}b_2^{2(n-1)}\cdots b_{n}^{2} = D_n\,,\quad D_n \equiv \det(\mu_{i+j})_{0\leq i,j\leq n}\,.
\end{align}
Using this, the Lanczos coefficients can be obtained from the moments as follows.
\begin{align}
b_n^2 = \frac{D_{n-2}D_n}{D_{n-1}^2} \quad (D_{-1}=1)
\label{eq:moment-to-Lanczos}
\end{align}
Other sequential algorithms for obtaining Lanczos coefficients from moments are also known.
We will use these methods in our numerical analyses.

In quantum many-body systems, it is conjectured that the Lanczos coefficients $b_n$ asymptotically grow at most linearly $b_n\sim \alpha n$.
This linear growth of the Lanczos coefficients corresponds to an exponential growth of Krylov complexity $C_{\rm K}(t)\sim e^{2\alpha t}$.
The growth exponent $\alpha$ is expected to give an upper bound on the OTOC exponent $\lambda_{\rm L}$ of the operator under consideration as $\lambda_{\rm L}\leq 2\alpha$.
Therefore, by examining the Krylov complexity, we can obtain constraints on the OTOC index.
Since the Krylov complexity can be determined by the information in the two-point function, examining its behavior is relatively easy compared to computing OTOC.

The asymptotic behavior of the Lanczos coefficients corresponds to the tail behavior of the power spectrum $\Phi(\omega)$ (the Fourier transform of the auto-correlation function)
\begin{align}
\Phi(\omega) \equiv \int_{-\infty}^\infty dt\,e^{-i\omega t}C(t)\,,
\label{eq:power spectrum}
\end{align}
where we normalize the two-point function as $C(0)=1$.
Many previous studies have investigated the relationships between the behavior of the Lanczos coefficients and that of the power spectrum. For example, when the power spectrum is continuous, the following are known:
\begin{itemize}
\item Linear growth of the Lanczos coefficients, $b_n = \alpha n$, corresponds to \cite{viswanath1994recursion, Parker:2018yvk}
\begin{align}
    \Phi(\omega) = \frac{\pi}{\alpha}{\rm sech}\left(\frac{\pi\omega}{2\alpha}\right)\,.
\label{eq:sech-power-spectrum}
\end{align}
More generally, asymptotic linear growth $b_n\sim \alpha n~(n\to\infty)$ corresponds to the exponential decay of the tail of the power spectrum,
\begin{align}
    \Phi(\omega) \sim e^{-\pi|\omega|/2\alpha}\,.
\end{align}
This can be translated as the existence of the poles of the auto-correlation function at $t=\pm\frac{i\pi}{2\alpha}$.

\item The saturation $b_n\to b~(n\to\infty)$ of the Lanczos coefficient corresponds to the fact that $\Phi(\omega)$ has non-zero value only at $[-2b,2b]$ \cite{viswanath1994recursion, Barbon:2019wsy}. Conversely, the saturation value of the Lanczos coefficients can be determined from the information in support of the power spectrum.
In particular, when the Lanczos coefficients are perfectly constant $b_n=b$, the moments are given as $\mu_{2n} = b^{2n}C_n$ by the Catalan number $C_n=\frac{(2n)!} {(n+1)!n!}$ \cite{Parker:2018yvk}.
The corresponding auto-correlation function is
\begin{align}
    C(t) = \frac{J_1(2bt)}{bt}\,,
\end{align}
where $J_n$ is the Bessel function of the first kind. Then, the power spectrum becomes 
\begin{align}
    \Phi(\omega) = \frac{\sqrt{4b^2-\omega^2}}{b^2}\,\theta(2b-|\omega|)\,,
\end{align}
which actually has value only at $[-2b,2b]$ and this is Wigner semicircle itself.

\item In some systems, the Lanczos coefficients show staggering. This is a situation in which the Lanczos coefficients $b_n$ behave in an oscillatory manner such that they appear to be on two separate curves, depending on whether $n$ is even or odd, rather than one smooth curve.
In \cite{Camargo:2022rnt} the following conditions on the power spectrum are proposed for the absence of staggering.
\begin{description}
    \item[(I)] $\Phi(\omega)$ is finite at $\omega=0$, i.e., $0<\Phi(\omega=0)<\infty$.\label{condI}
    \item[(II)] $\Phi'(\omega)$ is a continuous function of $\omega$ over the support of $\Phi(\omega)$.
\end{description}
\end{itemize}
On the other hand, when the power spectrum is discrete, the relationship between the power spectrum and the Lanczos coefficient is less well understood.
It is pointed out in \cite{viswanath1994recursion} that when the power spectrum is expressed as the sum of a finite number of delta function peaks, the Lanczos coefficients will eventually decay to zero and the Lanczos algorithm terminates in a finite number of steps. 
Although the behavior of Lanczos coefficients and Krylov complexity in systems with finite degrees of freedom has been explored by studies such as \cite{Barbon:2019wsy, Rabinovici:2020ryf, Rabinovici:2022beu}, a systematic understanding using the power spectrum has not yet been obtained.
In Sec.~\ref{sec:toy-model}, we try to give a systematic understanding of the relationships between the behavior of the Lanczos coefficients and that of the power spectrum.

A prior study of Krylov complexity relevant to our paper is the analysis in \cite{Tang:2023ocr}. Here, Krylov complexity for the operator $B$ is calculated in two-sided RMT as introduced in section \ref{subsec:RMTandDSSYK}.
More specifically, this analysis is done in GUE.
It was found that the Lanczos coefficients grow linearly at the initial stage and then saturate to a constant value.
This saturation can be related to the fact that the spectrum of the theory is a Wigner semicircle, which has a bounded support.\footnote{For Krylov complexity, the energy density $\rho(\omega)$ is not a directly relevant quantity because Krylov complexity is determined from the power spectrum $\Phi(\omega)$, the Fourier transform of the two-point function. 
This $\Phi(\omega)$ is usually different from the actual energy density of the theory. However, if the energy density $\rho(\omega)$ has a support, we can expect the behavior of the Lanczos coefficients and Krylov complexity at late stages. This point will be discussed later in Sec.~\ref{sec:spectrum-support}.}
This analysis is equivalent to computing the Krylov complexity of the fermion chain in the $q \to 0$ DSSYK model.
In Sec.~\ref{sec:Krylov-in-DSSYK}, we compute the Krylov complexity of fermion chain in DSSYK model for $q\to0$, and essentially we compute the same thing.
However, our analysis will move away from this limit and analyze the Krylov complexity of fermion chain in the DSSYK model in another limit. This means we will move away from simple GUE case on the RMT side.


\section{Krylov complexity of fermion chain in the DSSYK model}
\label{sec:Krylov-in-DSSYK}
\subsection{Krylov complexity of fermion chain}
From the moments of the two-point function \eqref{eq:2pt-analytic}, we calculate the Lanczos coefficients and analyze the Krylov complexity of the fermion chain operator. 
However, since its general expression is complicated, we mainly consider the limit which is easy to analyze. In the following, we set $\mathcal{J}=1$.

\subsubsection{In the case of $q, \tilde{q}\to0$}
In this case, only the chord diagram survives such that there is no intersection of the $H$ chord and the $M$ chord, as can be seen from the \eqref{eq:2pt-chord}. In other words, the diagram is completely divided by the $M$ chord and the two-point function factorizes to the product of the two partition functions:
\begin{align}
\langle
{\rm Tr}\,
\wick{
    \c1 M
    e^{-\beta_1H}
    \c1 M
    e^{-\beta_2H}
}
\rangle
\to \langle {\rm Tr}{\,e^{-\beta_1 H}}\rangle \langle {\rm Tr}{\,e^{-\beta_2 H}}\rangle \quad (\tilde{q}\to0)\,.
\label{eq:2pt-0th-order}
\end{align}
This is confirmed by the fact that the last factor of \eqref{eq:2pt-analytic} is 1 with $\tilde{q}\to0$, and the integral is completely separated and becomes the product of two partition functions.\footnote{Strictly speaking, $\tilde{q} \to 0$ limit yields the following. 
\begin{align}
\frac{(\tilde{q}^2;q)_\infty}{(\tilde{q}e^{i(\pm\theta_1\pm\theta_2)};q)_\infty}\sim 1 - 4 \cos \theta_1 \cos \theta_2 \frac{\tilde{q}^2}{1-q}+O(\tilde{q}^4)
\end{align}
Suppose the result of the $\theta$ integration with respect to the second term on the right-hand side is non-zero and finite. Then, the partition function factorizes in the range $\tilde{q}^2 \ll 1-q$.}

\begin{figure}[t]
\centering
    \includegraphics[width=90mm]{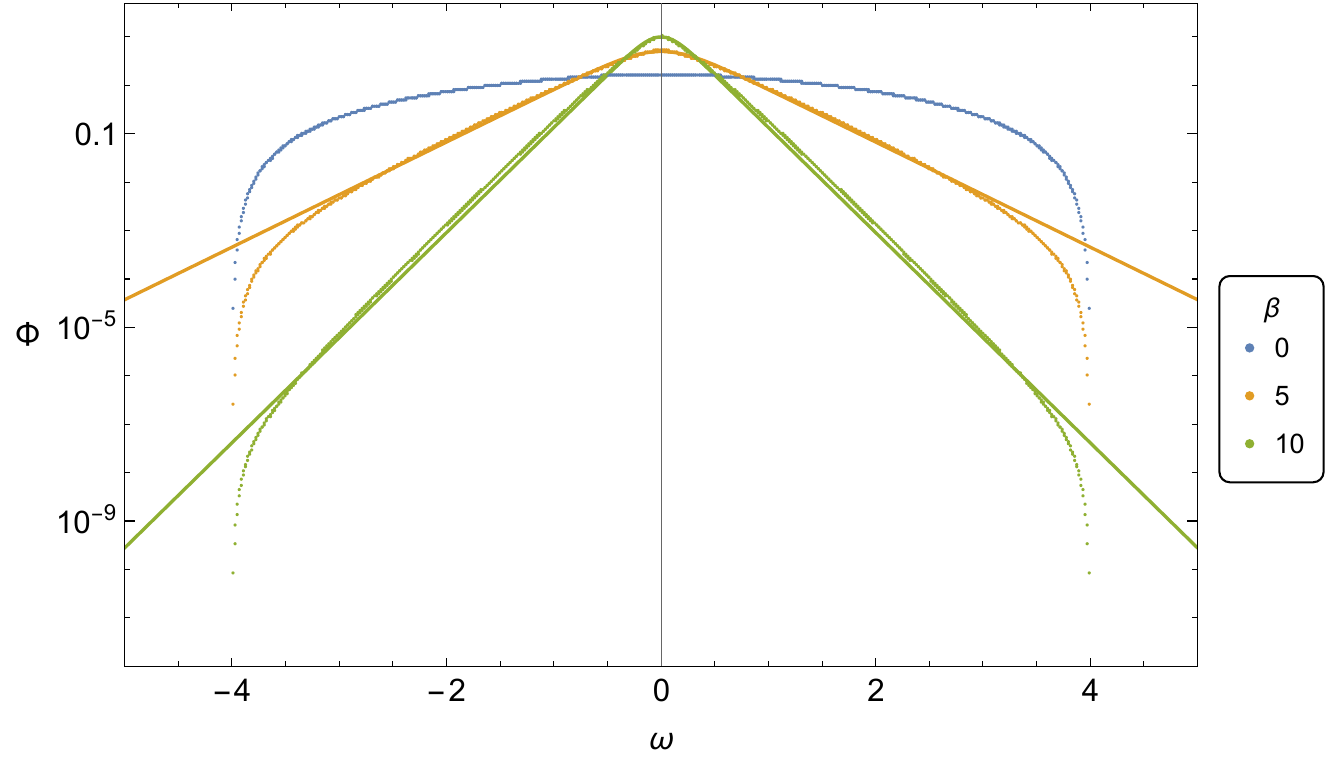}
    \caption{The $\omega$ dependence of the power spectrum of \eqref{eq:2pt-finite-temp}. The two-point function is normalized to have an initial value of unity. The solid line is $\beta\,{\rm sech}\left(\beta\omega/2\right)$.}
	\label{fig:q-to-0-power-spectrum}
\end{figure}

Now, if we also impose $q\to0$ to \eqref{eq:2pt-0th-order}, the partition function can be expressed using the deformed Bessel function as we have already seen, as in \eqref{eq:partition-function}, so \eqref{eq:2pt-0th-order} becomes 
\begin{align}
\langle
{\rm Tr}\,
\wick{
    \c1 M
    e^{-\beta_1H}
    \c1 M
    e^{-\beta_2H}
}
\rangle
\to
\frac{I_1(2\beta_1)}{\beta_1}\frac{I_1(2\beta_2)}{\beta_2}
\quad
(q,\tilde{q}\to0)\,.
\label{eq:2pt-finite-temp}
\end{align}
This leads back to the prior research of the Krylov complexity in RMT \cite{Tang:2023ocr}. Since the spectrum is continuous and bounded, the power spectrum of the two-point function is also continuous and bounded,\footnote{This point is discussed below in Sec.~\ref{sec:spectrum-support}.} and the Lanczos coefficient always settles to a constant value.
It has been confirmed that, in the same reference, depending on the finite temperature $\beta$, the Lanczos coefficients initially exhibit a linear increase $b_n \sim \alpha n$, and their slope can be well approximated by $\alpha = \frac{\pi}{\beta}$.
In Fig.~\ref{fig:q-to-0-power-spectrum}, we show the power spectrum \eqref{eq:power spectrum} of the two-point function \eqref{eq:2pt-finite-temp}.
The solid line is $\beta\,{\rm sech}\left(\beta\omega/2\right)$.
As we move to lower temperatures, the power spectrum behaves more like $\beta\,{\rm sech}\left(\beta\omega/2\right)$ over a wider $\omega$ range.
This is reflected in the longer linear increase of the Lanczos coefficient.
On the other hand, the power spectrum has a value only in the range $[-4,4]$, independent of temperature.
This implies that the asymptotic value of the Lanczos coefficient is constant regardless of temperature.

When $\tilde{q}\to0$, the OTOC of the fermion chain operator factorizes into smaller correlation functions and shows no exponential time dependence.
This OTOC behavior is similar to the results of previous studies in GUE random matrix theory \cite{Cotler:2017jue}.
In this case, the bound $\lambda_{\rm L}\leq 2\alpha$ of the OTOC exponent by the exponential growth rate $\alpha$ of the Krylov complexity is trivially satisfied.

\subsubsection{In the case of $q, \tilde{q}\to 1^-$ with $\tilde{q}=q^m$}
Now we consider the case where $q, \tilde{q}\to 1^-$ with $\tilde{q}=q^m$.
In terms of $\lambda$, this corresponds to $\lambda\to0$.
The form of the two-point function is examined in detail in \cite{Berkooz:2018qkz} and the results depend on the temperature regime considered.
The choice of $\tilde{q}=q^m$ with $m$ an integer corresponds to considering a fermion chain operator $M$ with $p'=mp$.\footnote{If $q$ and $\tilde{q}$ are taken independently and $q$ is left unchanged and $\tilde{q}\to1$, the two-point function becomes $\langle Z(\beta_1+\beta_2)\rangle$ which is the partition function itself $\langle Z(\beta)\rangle$ and is independent of time. Therefore, in this case, Krylov complexity of the fermion chain does not grow.}
In the low temperature, one expects that the conformal symmetry emerges and the dimension of $M$ becomes $m$.

\subsubsection*{Low temperature regime}
When $\lambda^{-3/2} \gg \beta\gg \lambda^{-1/2}$, and $t\ll \lambda^{-3/2}$,
the two-point function becomes \cite{Berkooz:2018qkz}
\begin{align}
    \langle
{\rm Tr}\,
\wick{
    \c1 M
    e^{-\beta_1H}
    \c1 M
    e^{-\beta_2H}
}
\rangle
\to (-i\partial_t)^{2m-2}\frac{1}{\cosh^2\left(\frac{\pi t}{\beta}\right)}
\label{eq:low-temperature}
\end{align}
up to numerical coefficients independent of $t$.
This two-point function is the same one used when calculating for Krylov complexity in a particular conformal field theory \cite{Dymarsky:2021bjq}.
They considered free massless scalar fields in general dimensions.
The conformal dimension $\Delta$ of the free scalar field corresponds to $m$ in \eqref{eq:low-temperature}.
According to their results, Krylov complexity from \eqref{eq:low-temperature} grows exponentially 
as $C_{\rm K}(t) \sim e^{2\alpha t}$ with $\alpha=\pi/\beta$.
In \cite{Berkooz:2018jqr}, the OTOC of the fermion chain operators is considered in the low-temperature regime, $\lambda^{3/2}\ll T \ll \lambda^{1/2}$, and small $\lambda\ll 1$.
The Lyapunov exponent is found as $\lambda_{\rm L} = 2\pi T -4\pi \lambda^{-1/2} T^2 + \cdots$, so the bound $\lambda_{\rm L} \leq 2\alpha$ holds.

Let us mention to what extent the approximation used in \eqref{eq:low-temperature} is valid and how Krylov complexity is expected to behave in very late time.
The expression \eqref{eq:low-temperature} of the two-point function is obtained by taking only leading order term.
For very late time, this approximation breaks down.
Therefore, what can be known from \eqref{eq:low-temperature} is the early time behavior of Krylov complexity, and the very late time behavior cannot be obtained from this approximation.
To understand the very late time behavior of Krylov complexity, which is outside the valid range of the approximation in \eqref{eq:low-temperature}, it is necessary to consider the full two-point function.
Although it is difficult to perform detailed calculations from the full two-point function specifically, the nature of the energy spectrum can provide information on the asymptotic behavior of the Lanczos coefficients.
When $\lambda \ll 1$, the range of the energy spectrum becomes $[-2/\sqrt{\lambda},2/\sqrt{\lambda}]$.\footnote{This comes from the small $\lambda$ approximation of maximum energy for DSSYK \eqref{eq:analytic-partition-function}.}
Although $\lambda\ll 1$, due to the low temperature condition of $\lambda^{3/2}\ll T \ll \lambda^{1/2}$, $\lambda$ must in turn satisfy $T^2\ll \lambda \ll T^{2/3}$ for a fixed temperature $T$.
Therefore, $\lambda$ takes a small but finite value, and the energy spectrum remains bounded.
Then, it follows that the power spectrum of the full two-point function will have compact support with the size of order $O(1/\sqrt{\lambda})$.
Note here that this does not depend on the degrees of freedom $N$.

The fact that the power spectrum has a compact support of size $O(1/\sqrt{\lambda})$ suggests that if the Lanczos coefficients are calculated from the full two-point function, the growth of the Lanczos coefficient saturates at $b=O(1/\sqrt{\lambda})$.
Then, as discussed later in Sec.~\ref{subsec:scrambling}, the exponential growth of Krylov complexity ends at the time $O(\beta\log\frac{\beta}{\sqrt{\lambda}})$ and it is expected to change to linear growth after that.
Notice that the time at which the change of the behavior occurs is independent of $N$, and in particular is smaller than the conventional fast scrambling time scale of $O(\log N)$. Also, again, it should be noted that this behavior of the Krylov complexity at very late time is the one expected when calculated from the full 2-point function, and cannot be seen from the approximate 2-point function \eqref{eq:low-temperature} obtained from taking the leading order.\footnote{
More concretely, \eqref{eq:low-temperature} is reliable up to $\beta = \tilde{\beta} \lambda^{-3/2}$ and $\tilde{\beta} \ll 1$. When we fix the temperature as $\beta = O(\lambda^{-3/2})$, we can show that the scrambling time is much bigger than the maximum time when \eqref{eq:low-temperature} is valid. Therefore, this low temperature approximation gets broken by the scarambling time.}



\subsubsection*{Very low temperature regime}
Now we consider $\beta \gg \lambda^{-3/2}$.
According to \cite{Berkooz:2018qkz}, the two-point function is found to be
\begin{align}
\langle
{\rm Tr}\,
\wick{
    \c1 M
    e^{-\beta_1H}
    \c1 M
    e^{-\beta_2H}
}
\rangle
\to (-i\partial_t)^{2m-2}\frac{1}{\left(1+\frac{4t^2}{\beta^2}\right)^{3/2}}
\label{eq:2pt-finite-temp-q-to-1}
\end{align}
up to numerical coefficients independent of $t$.
Using this expression of the two-point function, we study $\beta$ and $m$ dependence of the Lanczos coefficients and Krylov complexity in the following.


\begin{figure}[t]
    \begin{minipage}[t]{0.45\hsize}
        \centering
        \includegraphics[width=.9\columnwidth]{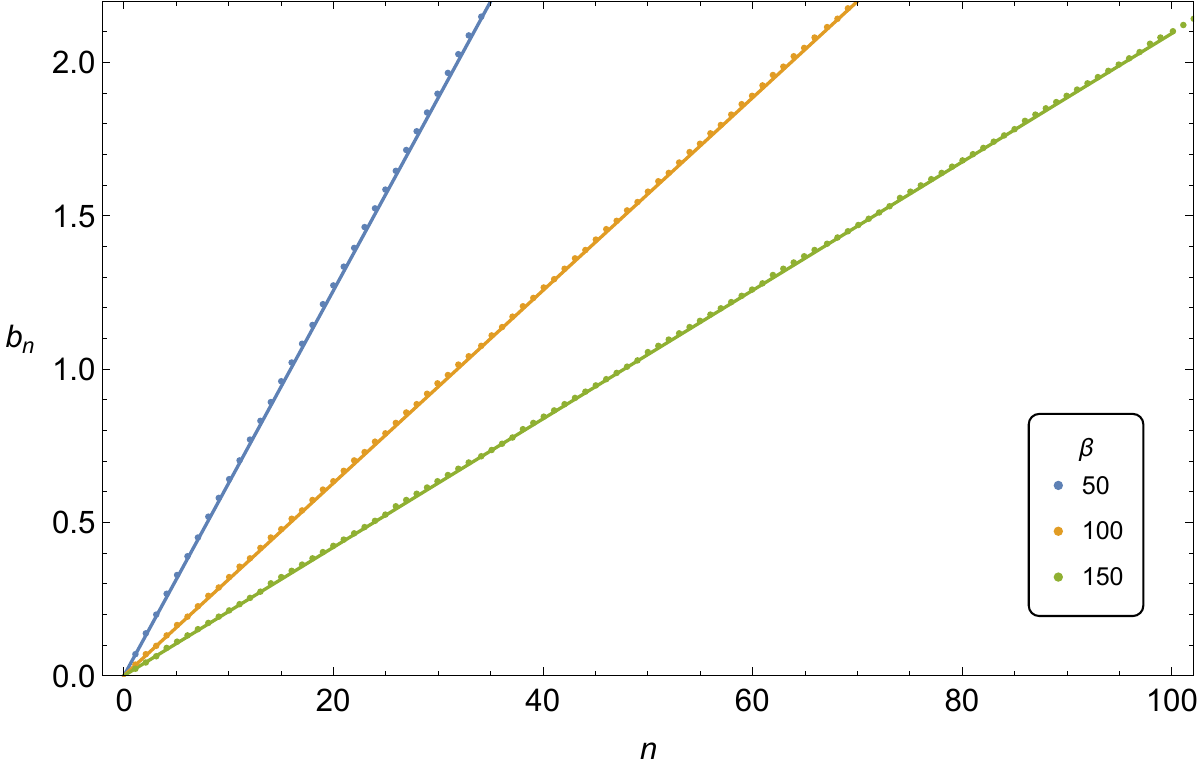}
        \subcaption{The Lanczos coefficients $b_n$ for several temperatures.}
        \label{fig:Lanczos-temp-dep}
    \end{minipage}
    \hspace{5mm}
    \begin{minipage}[t]{0.45\hsize}
        \centering
        \includegraphics[width=.95\columnwidth]{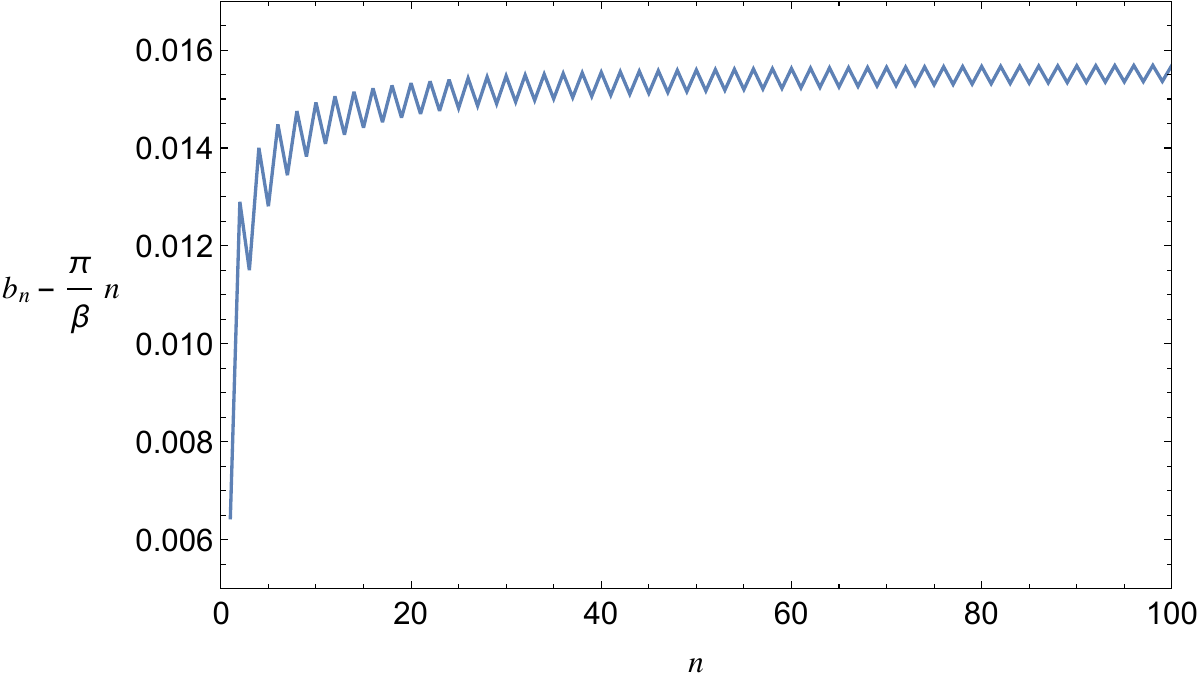}
        \subcaption{The difference between $b_n$ and $\frac{\pi}{\beta}n$ when $\beta=50$.}
        \label{fig:Lanczos-comp-with-linear}
    \end{minipage} \\
\caption{(a) The Lanczos coefficients computed from the two-point function \eqref{eq:2pt-finite-temp-q-to-1} with $m=1$. They appear to increase linearly already at the beginning. The solid line is $b_n=\frac{\pi}{\beta}n$.
(b) The difference between $b_n$ and $\frac{\pi}{\beta}n$ when $\beta=50$ is plotted, which confirms the staggering of the Lanczos coefficient.}
\label{fig:m=1-Lanczos}
\end{figure}

\begin{figure}[t]
    \begin{minipage}[t]{0.45\hsize}
        \centering
        \includegraphics[width=.95\columnwidth]{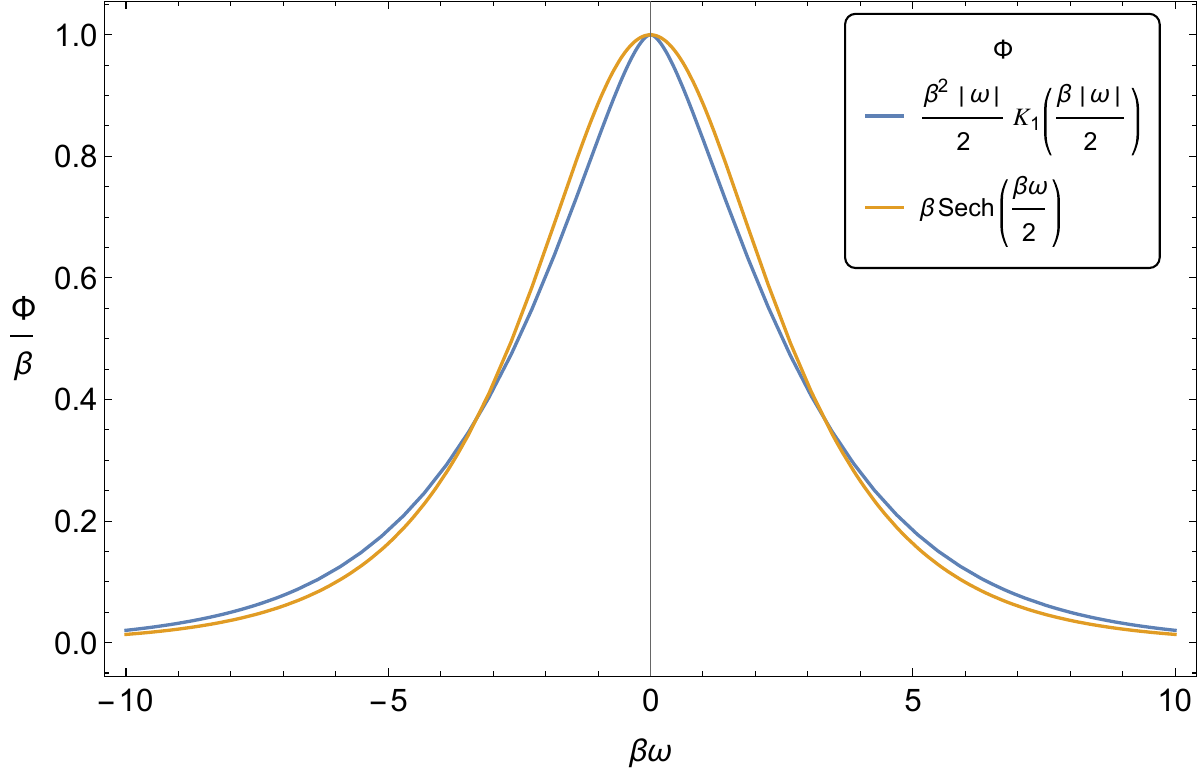}
        \subcaption{The power spectrum as a function of $\beta\omega$.}
        \label{fig:zeroth}
    \end{minipage}
    \hspace{5mm}
    \begin{minipage}[t]{0.45\hsize}
        \centering
        \includegraphics[width=.95\columnwidth]{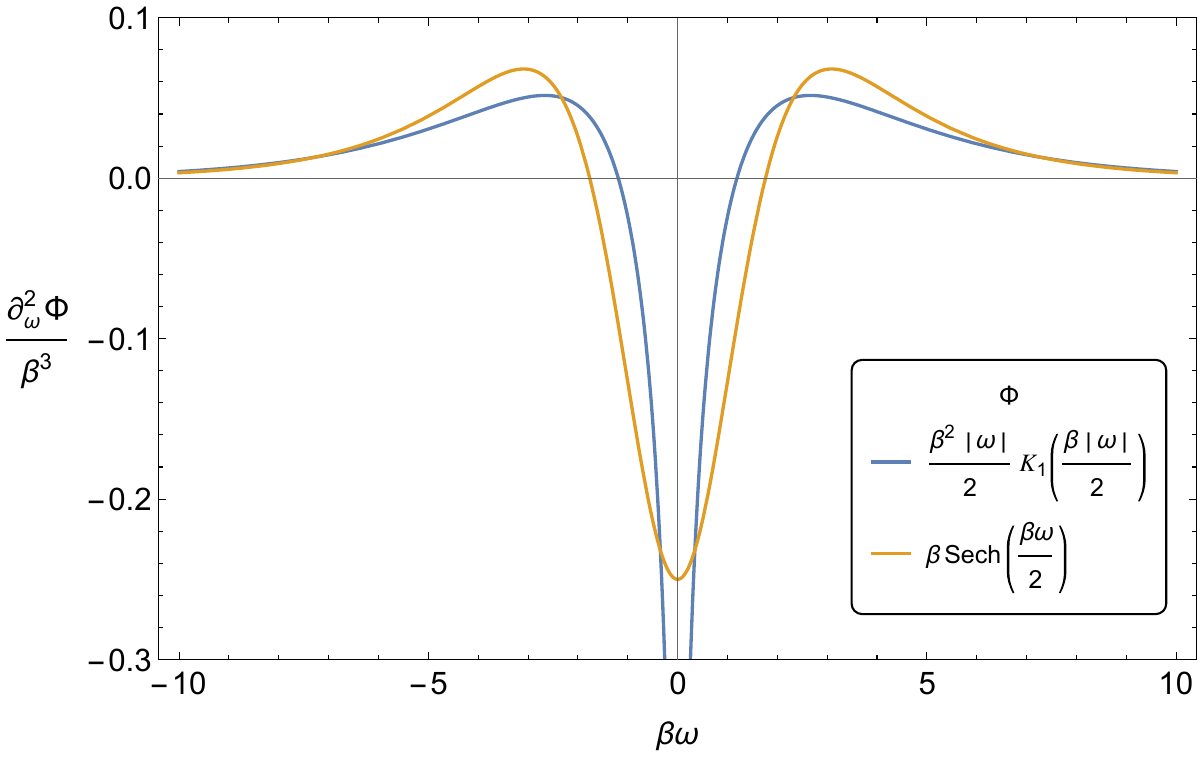}
        \subcaption{The second derivative of the power spectrum as a function of $\beta\omega$.}
        \label{fig:second}
    \end{minipage} \\
\caption{Comparison of \eqref{eq:power-spectrum-for-m=1} with $\beta\,{\rm sech}(\beta \omega/2)$. The second derivative of \eqref{eq:power-spectrum-for-m=1} diverges around $\omega=0$ as $\log(\beta\omega)$.}
\label{fig:q-to-1-power-spectrum}
\end{figure}

To begin with, let us examine the temperature dependence of Krylov complexity when $m=1$.\footnote{In this case, the time dependence of the two-point function is mathematically equivalent to that in the low-temperature limit when $\tilde{q}\to 0$ with $q\to0$ or $q\to1$.}
In Fig.~\ref{fig:Lanczos-temp-dep}, we show the Lanczos coefficients for several temperatures.
They appear to increase linearly already at the beginning.
However, closer examination reveals a slight staggering, i.e., the Lanczos coefficients $b_n$ appear to be on two different curves depending on whether $n$ is even or odd, as shown in Fig.~\ref{fig:Lanczos-comp-with-linear}.
The power spectrum of the two-point function we are dealing with here is
\begin{align}
    \Phi(\omega) = \frac{\beta^2|\omega|}{2}K_1\left(\frac{\beta|\omega|}{2}\right)\,,
\label{eq:power-spectrum-for-m=1}
\end{align}
where $K_1$ is the modified Bessel function.
Figure \ref{fig:q-to-1-power-spectrum} compares this with the power spectrum $\beta\,{\rm sech}(\beta \omega/2)$ when $b_n = \frac{\pi}{\beta} n$.
These two look very similar, but not the same.
Notice that the power spectrum \eqref{eq:power-spectrum-for-m=1} is finite at $\omega=0$, and the derivative $\Phi'(\omega)$ is continuous.\footnote{For $x>0$, $x K_1(x) = x\ln(x/2) I_1(x)+(\text{regular terms})$, where $I_1(x)=\frac{x}{2}\sum_{k=0}^\infty \frac{1}{k!(k+1)!}\left(\frac{x}{2}\right)^{2k}$.}
In \cite{Camargo:2022rnt} these properties were proposed to be the conditions for the absence of staggering.
The fact that \eqref{eq:power-spectrum-for-m=1} led to staggering means that stronger conditions are needed for the absence of staggering.
Since the second derivative of \eqref{eq:power-spectrum-for-m=1} diverges at $\omega =0$, one possible modification is to strengthen the condition on derivatives.
If instead of requiring that the first derivative be continuous, we require up to the continuity of the second derivative, then \eqref{eq:power-spectrum-for-m=1} would be out of condition and staggering would be allowed.
However, this is just a conjecture, and there could be other causes for the appearance of staggering. 

\begin{figure}[t]
\begin{minipage}[t]{0.45\hsize}
        \centering
        \includegraphics[width=.9\columnwidth]{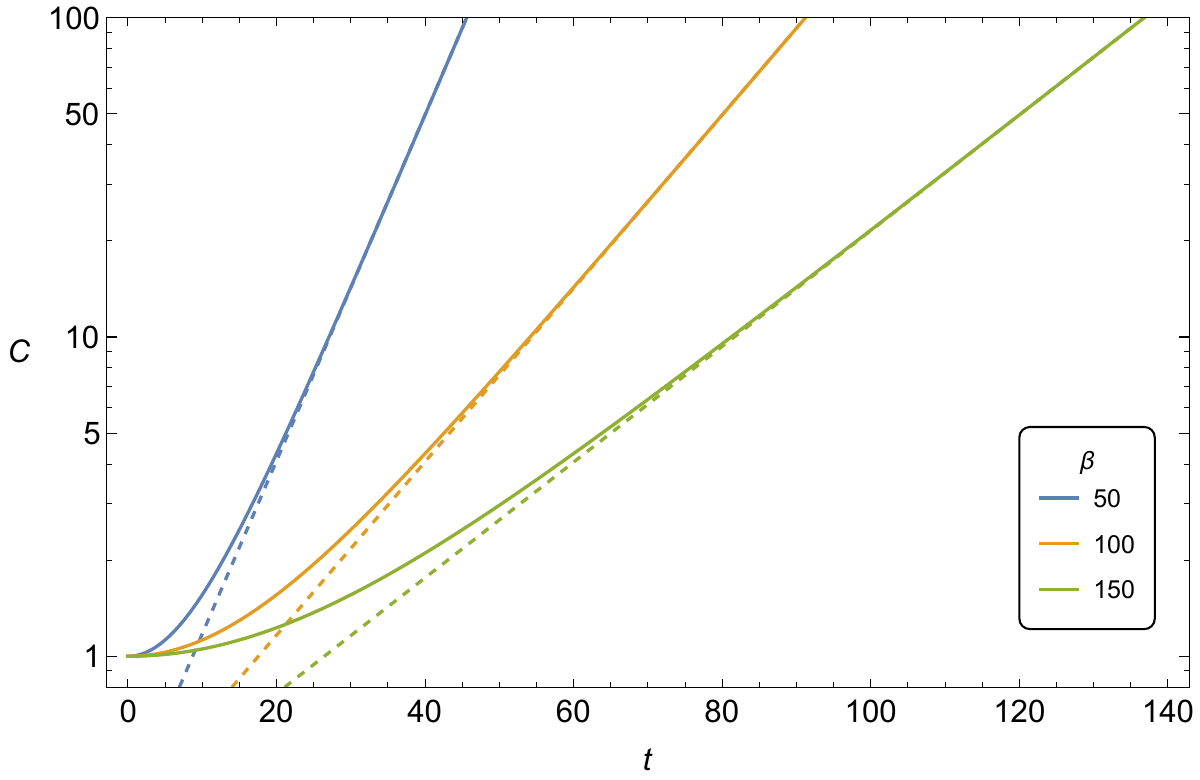}
        \subcaption{The time dependence of Krylov complexity}
        \label{fig:Krylov-temp-dep}
      \end{minipage}
      \hspace{5mm}
      \begin{minipage}[t]{0.45\hsize}
        \centering
        \includegraphics[width=.95\columnwidth]{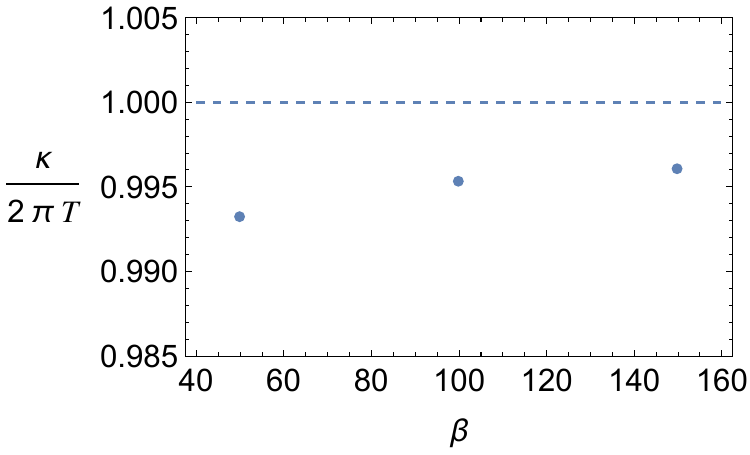}
        \subcaption{The temperature dependence of the growth exponent of Krylov complexity}
        \label{fig:Krylov-exponent}
      \end{minipage} \\
    \caption{(a) The time dependence of the Krylov complexity computed from the Lanczos coefficients in Fig.~\ref{fig:Lanczos-temp-dep}. The asymptotic exponential increase can be seen. (b) is the growth exponent $\kappa$ of Krylov complexity ($C_{\rm K}(t)\sim e^{\kappa t}$) found by numerical fitting. The values are close to the upper bound $\kappa \leq 2\pi T$ .}
     \label{fig:m=1-Krylov}
\end{figure}

\begin{figure}[t]
\begin{minipage}[t]{0.45\hsize}
        \centering
    \includegraphics[width=.9\columnwidth]{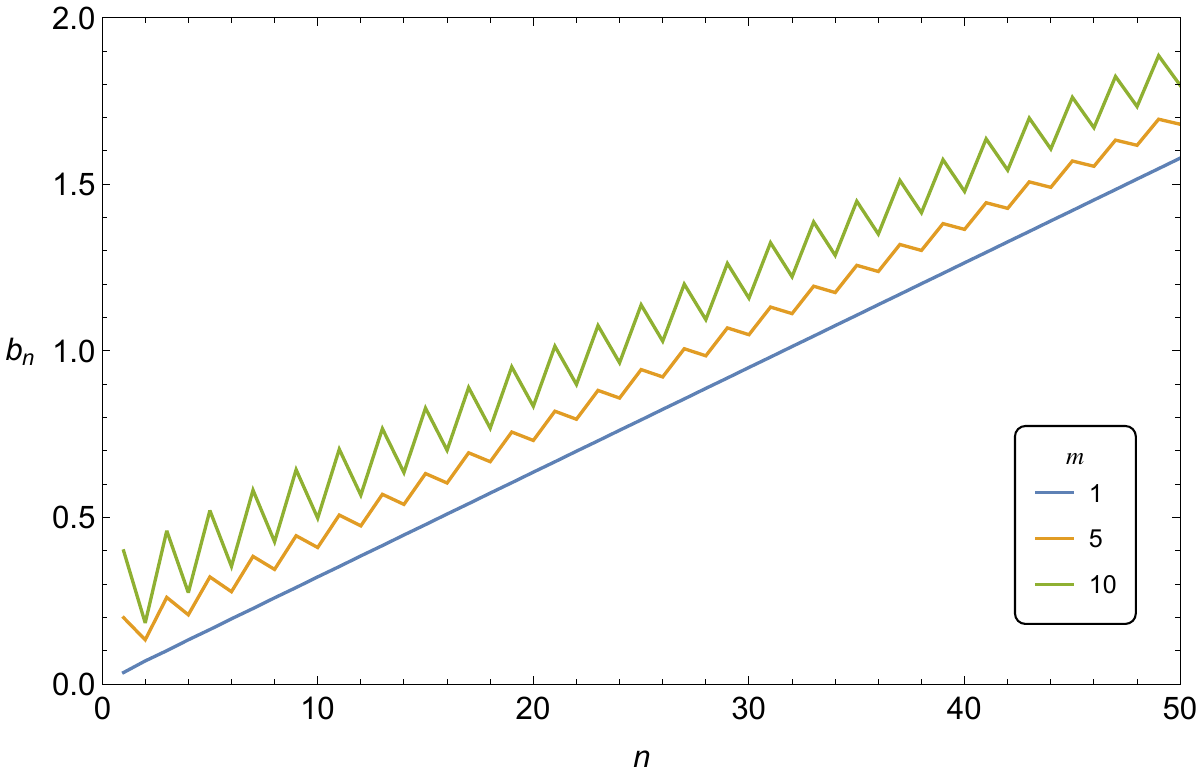}
        \subcaption{The Lanczos coefficients $b_n$ for several $m$.}
        \label{fig:Lanczos-m-dep}
      \end{minipage}
      \hspace{5mm}
      \begin{minipage}[t]{0.45\hsize}
        \centering
    \includegraphics[width=.95\columnwidth]{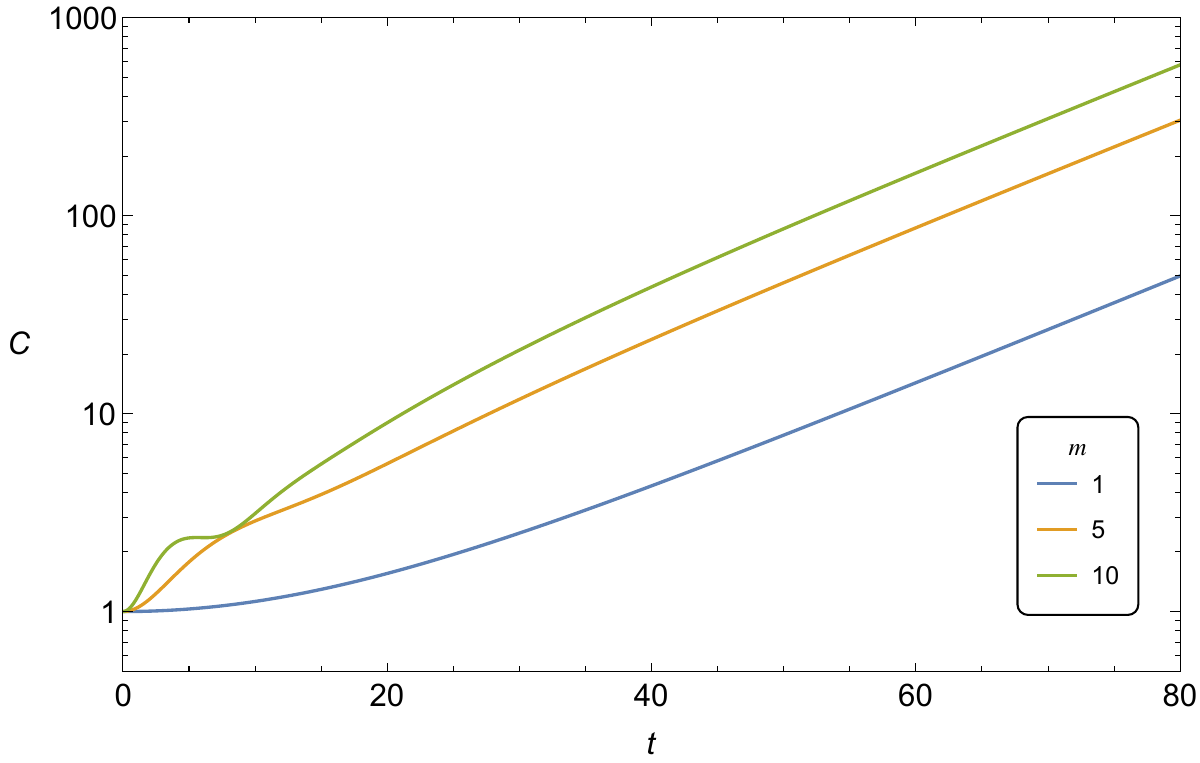}
        \subcaption{The time dependence of Krylov complexity for several $m$.}
        \label{fig:Krylov-m-dep}
      \end{minipage} \\
    \caption{The Lanczos coefficients and the time dependence of Krylov complexity for $\beta=100$. (a) As $m$ increases, the staggering of the Lanczos coefficients becomes more apparent. When the Lanczos coefficient reaches about $O(1/\sqrt{\lambda})$ (which is smaller than $O(\beta^{1/3})$), the growth should terminate and approach a constant value. (b) The initial behavior of the Krylov complexity changes accordingly. As $m$ becomes larger, Krylov complexity also becomes larger.}
     \label{fig:Lanczos-Krylov-m-dep}
\end{figure}

Figure \ref{fig:Krylov-temp-dep} shows the time dependence of Krylov complexity.
The exponential growth can be seen clearly.
However, note that, exactly as in the low temperature case, \eqref{eq:2pt-finite-temp-q-to-1} is an approximate expression obtained by taking only the leading contribution in $\beta \gg \lambda^{-3/2}$.
This approximation is invalid in very late time.\footnote{This approximation is expected to be valid up to $t = O(\lambda^{-3/2})$. This is bounded by the scrambling time $O(\beta \log \frac{\beta}{\sqrt{\lambda}})$.}
To know the very late time behavior of the Krylov complexity in detail, we need to use the full two-point function.
By exactly the same arguments as in the low temperature case, we can deduce that the power spectrum computed from the full two-point function has compact support of size $O(1/\sqrt{\lambda})$.
Then, the Lanczos coefficient reaches plateau at $O(1/\sqrt{\lambda})$, which is smaller than $O(\beta^{1/3})$ in the very low temperature region $\beta\gg\lambda^{-3/2}$.
Correspondingly, the exponential growth in Krylov complexity is expected to end at time $O(\beta\log\frac{\beta}{\sqrt{\lambda}})$ as discussed later in Sec.~\ref{subsec:scrambling}. 
After that the behavior of Krylov complexity turns into a linear growth.
Notice that this time scale can be larger than $O(\beta\log\frac{1}{\lambda})$ when $\beta \gg \lambda^{-3/2}$.
When $\lambda\ll 1$, this is also much larger than $O(\beta)$. Therefore, the behavior of the Krylov complexity is reliable within the scope of Fig.~\ref{fig:Krylov-temp-dep}. \footnote{Strictly speaking, this very low temperature limit is valid for $\beta$ more than $\tilde{\beta} \lambda^{-3/2}$ and $\tilde{\beta} \gg 1$. In Fig.~\ref{fig:m=1-Krylov} and Fig.~\ref{fig:Lanczos-Krylov-m-dep}, we consider fixing $\beta$ on the order of $\lambda^{-3/2}$ in order to make the plotting range include enough time region to approximate a two-point function. For example, for $\lambda = 0.1$, $\lambda^{-3/2} \sim 30$ and $\beta \sim 3 \times \lambda^{-3/2}$.}
The change from exponential increase to linear increase occurs at a much later time.
In Fig.~\ref{fig:Krylov-exponent} is shown the growth exponent $\kappa$ of Krylov complexity ($C_{\rm K}(t)\sim e^{\kappa t}$) found by numerical fitting.
They are close to the upper bound $\kappa \leq 2\pi T$.
The slightly smaller value is due to fitting over a finite time range, and in the limit of late time, it is expected to saturate the upper bound.

\begin{figure}[t]
\centering
	\includegraphics[width=90mm]{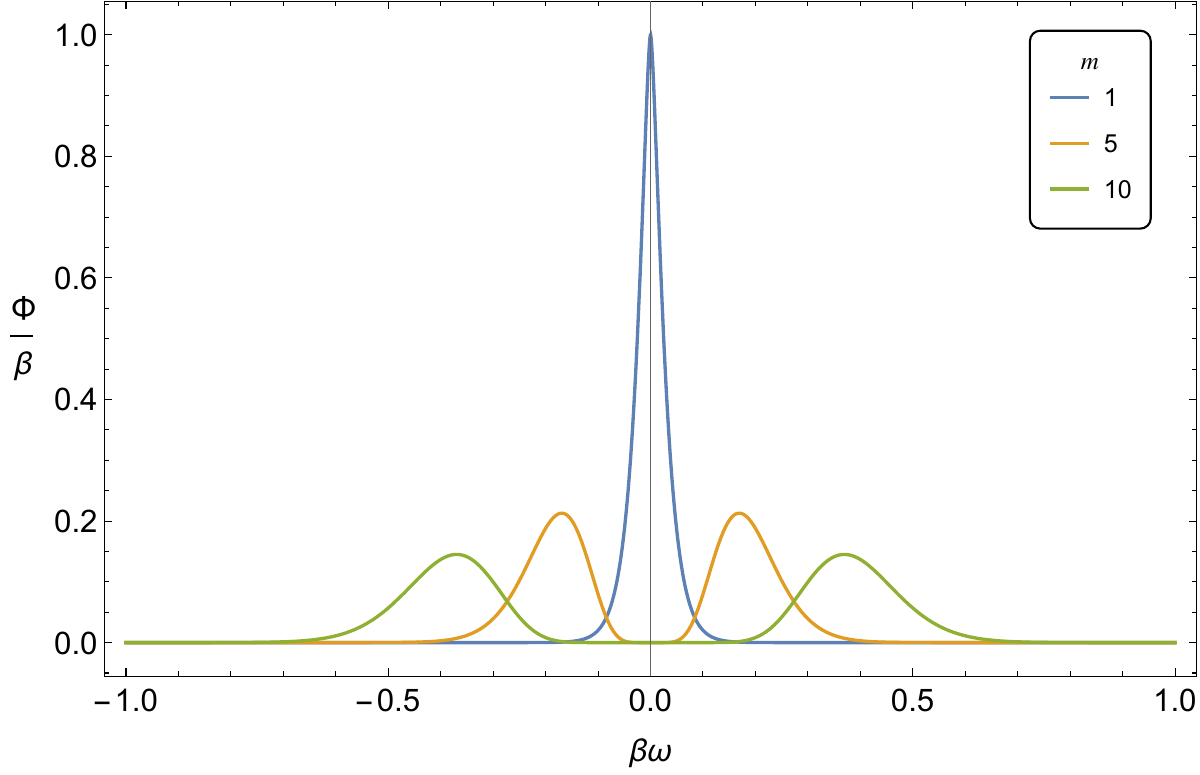}
	\caption{The $\omega$ dependence of the power spectrum for several $m$. The power spectrum for $m$ in general is the power spectrum for $m=1$ multiplied by $\omega^{2m-2}$, and as $m$ increases, the value around $\omega=0$ becomes smaller. Therefore, the power spectrum consists of two peaks, and as $m$ increases, the peaks move away from each other.}
	\label{fig:q-to-1-power-spectrum-m-dep}
\end{figure}

Now we fix $\beta = 100$ and look at the change in Lanczos coefficients and Krylov complexity when $m$ is changed.
Recall that, before taking the double scaling limit, $m$ is related to the length $p'$ of the fermion chain operator as $p'=mp$, where $p$ is the length of the Hamiltonian.
Therefore, changing $m$ might be regarded as changing the size of the fermion chain $M$.
Figure \ref{fig:Lanczos-Krylov-m-dep} shows the results for the Lanczos coefficients and Krylov complexity.
The staggering of the Lanczos coefficients increases as $m$ is increased.
This is likely to correspond to the fact that the power spectrum consists of two peaks, as shown in Fig.~\ref{fig:q-to-1-power-spectrum-m-dep},
but there is currently no analytical method that specifically links this power spectrum behavior to the Lanczos coefficient.
More detailed methods will need to be developed in the future to specifically confirm this.
Note that, in the current case, the power spectrum breaks the first condition ({\bf I}), in page \pageref{condI}, for the absence of staggering in \cite{Camargo:2022rnt}.
Although the Lanczos coefficients are staggering, overall they increase linearly with $n$ and the width of the staggering becomes smaller.
As can be seen in Fig.~\ref{fig:Krylov-m-dep}, Krylov complexity grows exponentially at late times.
It can also be seen that Krylov complexity takes a larger value for larger $m$.
A larger $m$ means a longer fermion chain $M$, and Fig.~\ref{fig:Krylov-m-dep} implies that a longer fermion chain earns more complexity.
Although small differences in the staggering  behavior of Lanczos coefficients can be seen, the Krylov complexity are very similar to those calculated for the free scalar field case in \cite{Dymarsky:2021bjq} mentioned before.
We can see that the conformal dimension $\Delta$ of the scalar field and $m$ correspond to each other. 

\subsection{Scrambling time}
\label{subsec:scrambling}
Let us comment on the scrambling time.
The definition of scrambling time is ambiguous and varies slightly in the literature.
In the context of Krylov complexity, the scrambling time is sometimes defined as the time when the value of Krylov complexity is of the order of the number of degrees of freedom of the system.
However, the actual choice of this order is not very standardized in the literature.
Here, we consider the time $t_*$ at which the Krylov complexity changes from exponentially increasing to linearly increasing as the scrambling time \cite{Barbon:2019wsy, Tang:2023ocr}.

This $t_*$ can be roughly evaluated using the value of $n=n_*$ around which the behavior of the Lanczos coefficients changes from a linear growth $b_n\sim \alpha n$ to a plateau $b_n\sim b$.
For this purpose, we note that the Krylov complexity can be regarded as the expectation value of the site location of a one-dimensional chain system with Lanczos coefficients as hopping.
Before the scrambling time $t\leq t_*$, Krylov complexity increases exponentially as $e^{2\alpha t}$, and this exponential diffusion on the one-dimensional chain continues until near the $n=n_*$ site, so the scrambling time is $e^{2\alpha t_*} \sim n _*$, from which we find $t_*\sim \alpha^{-1}\log n_*$.
Since $n_*\sim b/\alpha$ from the behavior of the Lanczos coefficient, we obtain $t_*\sim \alpha^{-1}\log (b/\alpha)$.
The plateau value $b$ of the Lanczos coefficients is determined by the size of the support of the power spectrum $\Phi(\omega)$.
This is of the same order as the maximum energy eigenvalue of the system under consideration, from which we can estimate the dependence of $t_*$ on the number $N$ of degrees of freedom.

In the case of the DSSYK model, the plateau value of the Lanczos coefficient becomes $b\sim \mathcal{J}$ in the $q\to 0$ limit and $b\sim \mathcal{J}/\sqrt{\lambda}$ in the $q\to 1^-$ limit.
Since neither of these depends on the number $N$ of Majorana fermions, the scrambling time $t_*$ also does not depend on $N$. In other words, in this sense, the DSSYK model is {\it hyperfast scrambling}.
Although our specific analysis was carried out under a particular limit of $q,\tilde{q}$, the fact that scarmbling does not depend on the degrees of freedom $N$ can be expected to hold in general.  
This is because the power spectrum of the DSSYK model is bounded by $N$-independent values from the claim introduced in Sec.~\ref{sec:spectrum-support}.
In the case of the usual SYK model, since the maximum energy eigenvalue increases as a power function of $N$, the scrambling time becomes $t_*\sim \alpha^{-1} \log N$, which is consistent with the conventional fast scrambling property.

In the above, we call a system whose scrambling time is $O(N^0)$ {\it hyperfast scrambling}. Strictly speaking, in addition to this, the Lyapunov exponent $\lambda_{\rm L}$ determined from OTOC must be non-zero (or at least a interacting theory like the SYK model). This is because the Krylov complexity gives only an upper bound on $\lambda_{\rm L}$, so the scrambling time that Krylov complexity determines does not necessarily mean that of the OTOC. In fact, there are free systems where the Lanczos coefficients behave similarly (increasing linearly at the beginning and saturating) and have a scrambling time of the same order (e.g. XY model \cite{Avdoshkin:2022xuw,Parker:2018yvk}), but these are clearly not chaotic systems.
This fact that Krylov complexity can grow exponentially even in free theory suggests that scrambling in the sense of Krylov complexity is different from scrambling as a diffusion of perturbations in real space.
It is more accurate to say that we are thinking of scrambling in operator space \cite{Parker:2018yvk,Tang:2023ocr}.


\section{Characterization by power spectrum}
\label{sec:toy-model}
As we have mentioned, if the support of the power spectrum is bounded, the growth in the Lanczos coefficient will terminate somewhere. Correspondingly, the late time behavior of the Krylov complexity would be at most linear growth.
In such systems, we should focus on the early time region in examining the characteristic behavior of each operator.
In this section, we investigate the behavior of Lanczos coefficients from the viewpoint of the power spectrum using artificial power spectrum.

\subsection{Toy power spectrum}
In the case of finite degrees of freedom or bounded quantum systems, the energy spectrum is discrete.
To investigate more systematically the differences in the behavior of Lanczos coefficients in discrete and continuous systems, we consider the following toy power spectrum:
\begin{align}
    \Phi(\omega) = \mathcal{N}\sum_{l=-L}^L{\rm sech}\left(\frac{\pi\omega_l}{2\alpha}\right)\delta(\omega-\omega_l)\,, \quad \omega_l=\frac{l}{L}\omega_{\rm max}\,,
\label{eq:toy-power-spectrum}
\end{align}
where $\mathcal{N}$ is the normalization constant for $\int\frac{d\omega}{2\pi}\Phi(\omega)=1$.
The power spectrum \eqref{eq:toy-power-spectrum} has its support in $[-\omega_{\rm max},\omega_{\rm max}]$ and equally spaced delta function peaks of width $\Delta\omega=\omega_{\rm max}/L$, and their weights are given by ${\rm sech}(\pi\omega/2\alpha)$.\footnote{A discrete power spectrum with other weights is considered, for example, in \cite{Anegawa:2024wov}. In this reference, the case where $\omega_{\rm max}$ is very large and there is no bound in the power spectrum is discussed.} By analogy with the Boltzmann factor, in effect $\alpha$ also acts as a temperature.
In the following, we study the behaviors of Lanczos coefficients and Krylov complexity while varying the parameters $\Delta\omega, \omega_{\rm max}$ and $\alpha$.

\subsubsection{Varying $\Delta\omega$}
\begin{figure}[t]
      \begin{minipage}[t]{0.45\hsize}
        \centering
        \includegraphics[width=.95\columnwidth]{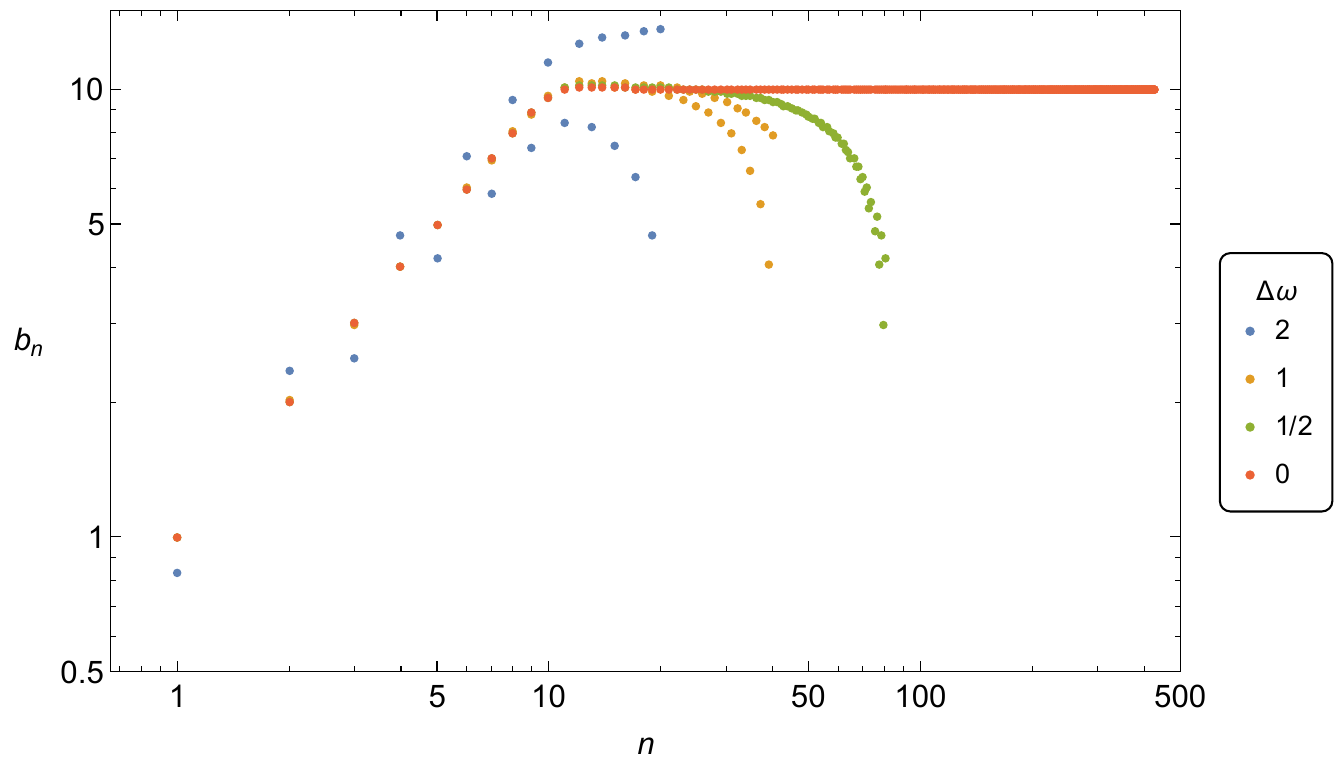}
        \subcaption{Lanczos coefficients for various $\Delta\omega$.}
        \label{fig:delta-omega-Lanczos}
      \end{minipage}
      \hspace{5mm}
      \begin{minipage}[t]{0.45\hsize}
        \centering
        \includegraphics[width=.95\columnwidth]{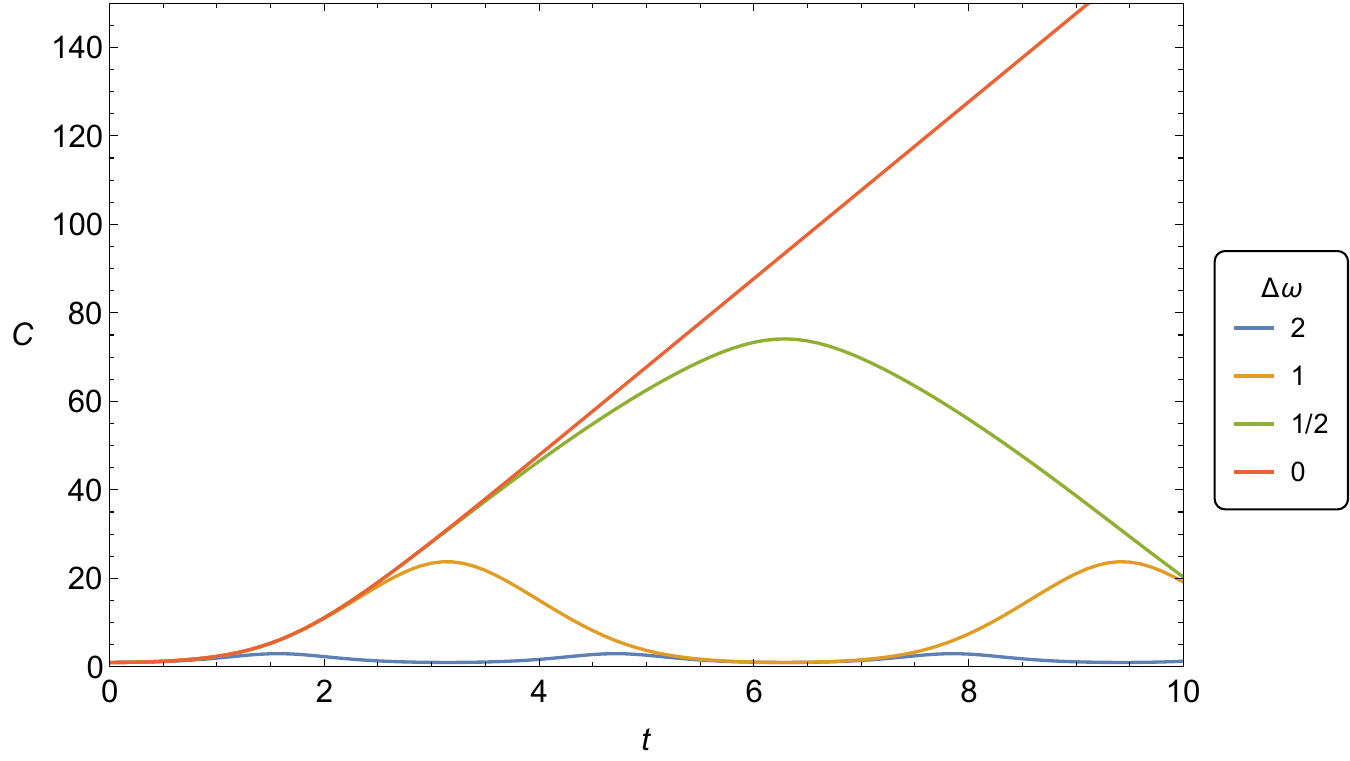}
        \subcaption{Krylov complexity for various $\Delta\omega$.}
      
        \label{fig:delta-omega-Krylov}
      \end{minipage} \\
    \caption{The behaviors of Lanczos coefficients and Krylov complexity when varying $\Delta\omega$ with $\omega_{\rm max}=20$ and $\alpha=1$.}
\end{figure}

To begin with, we fix $\omega_{\rm max}$ and $\alpha$ and vary $\Delta\omega$.
In Fig.~\ref{fig:delta-omega-Lanczos}, we show the Lanczos coefficients for various values of $\Delta\omega$ with $\omega_{\rm max}=20$ and $\alpha=1$.
As for the initial behavior of the Lanczos coefficients, when 
$\Delta\omega / \alpha \gtrsim O(1)$, the Lanczos coefficients show staggering. 
On the other hand, as $\Delta\omega /\alpha \ll 1$, the staggering becomes smaller. 
When $\Delta\omega / \alpha \gtrsim O(1)$, discreteness clearly appears, and $\Phi(\omega=0)$ is not finite. 
Therefore, both of the conditions for the absence of staggering proposed in \cite{Camargo:2022rnt} are indeed violated.
The behaviour of the Krylov complexity with the above change in the interval of the discrete spectrum is consistent with similar analyses \cite{Avdoshkin:2022xuw,Anegawa:2024wov}.

Next, we turn our attention to the remaining part of the Lanczos coefficients.
As we decrease $\Delta\omega / \omega_{\rm max}$, the dimension of the Krylov subspace should increase, and the Lanczos sequence should also become longer.
Indeed, Fig.~\ref{fig:delta-omega-Lanczos} is in line with the above expectation.
As $\Delta\omega / \omega_{\rm max}$ is decreased, the plateau of the Lanczos coefficients becomes longer, and in the continuous limit $\Delta\omega / \omega_{\rm max} \ll 1$, the plateau continues forever.
The plateau value of the Lanczos coefficient is determined by $\omega_{\rm max}$ and does not depend on $\Delta\omega$.
As is well known, the plateau of the Lanczos coefficient corresponds to a linear growth in Krylov complexity.
Figure \ref{fig:delta-omega-Krylov} shows the time dependence of Krylov complexity for various values of $\Delta\omega$.
The range of linear growth in Krylov complexity also becomes longer as $\Delta\omega$ is decreased. 
When $\Delta\omega / \omega_{\rm max} \gtrsim O(1)$, 
the Krylov subspace is small and recursions occur frequently, but as $\Delta\omega / \omega_{\rm max} \ll 1$,
the Krylov subspace becomes larger and recursions are less likely to occur.

\begin{figure}[t]
      \begin{minipage}[t]{0.45\hsize}
        \centering
        \includegraphics[width=.95\columnwidth]{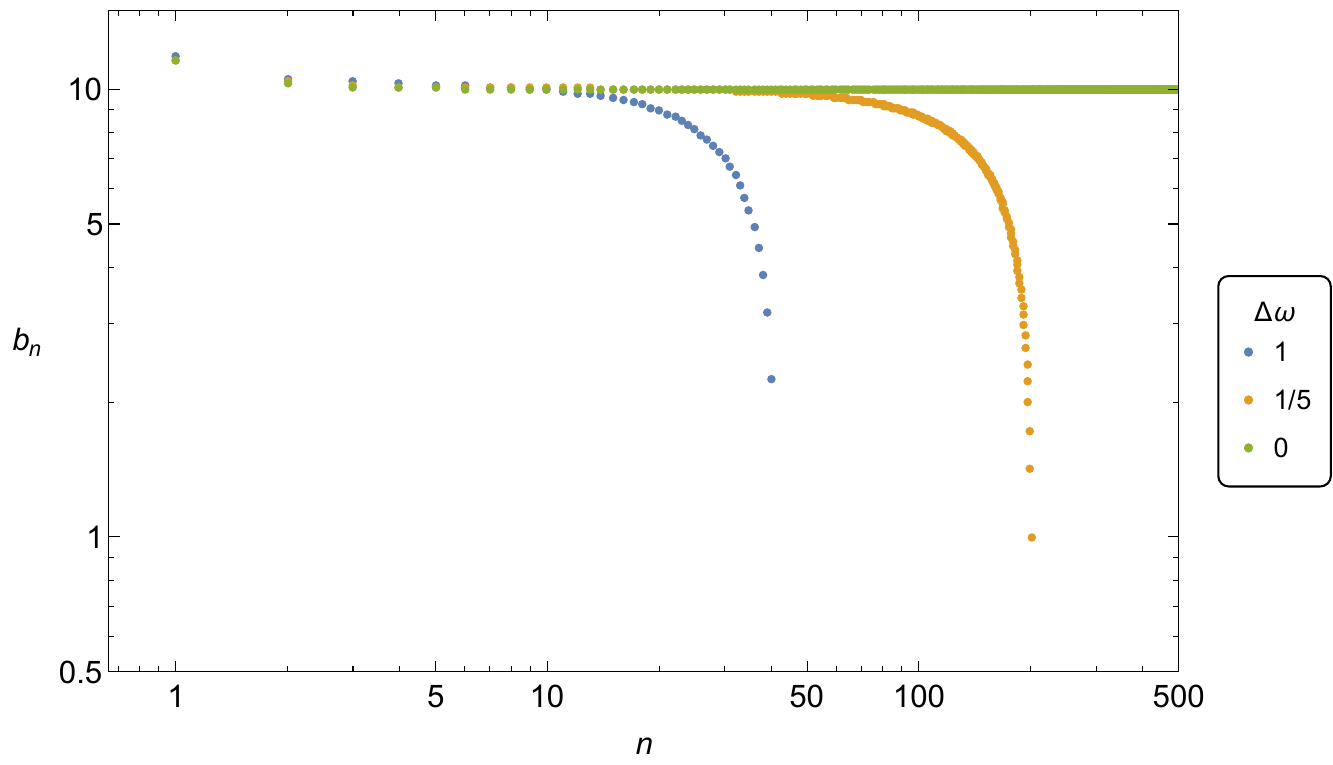}
        \subcaption{Lanczos coefficients for various $\Delta\omega$.}
        \label{fig:rectangular-Lanczos}
      \end{minipage}
      \hspace{5mm}
      \begin{minipage}[t]{0.45\hsize}
        \centering
        \includegraphics[width=.95\columnwidth]{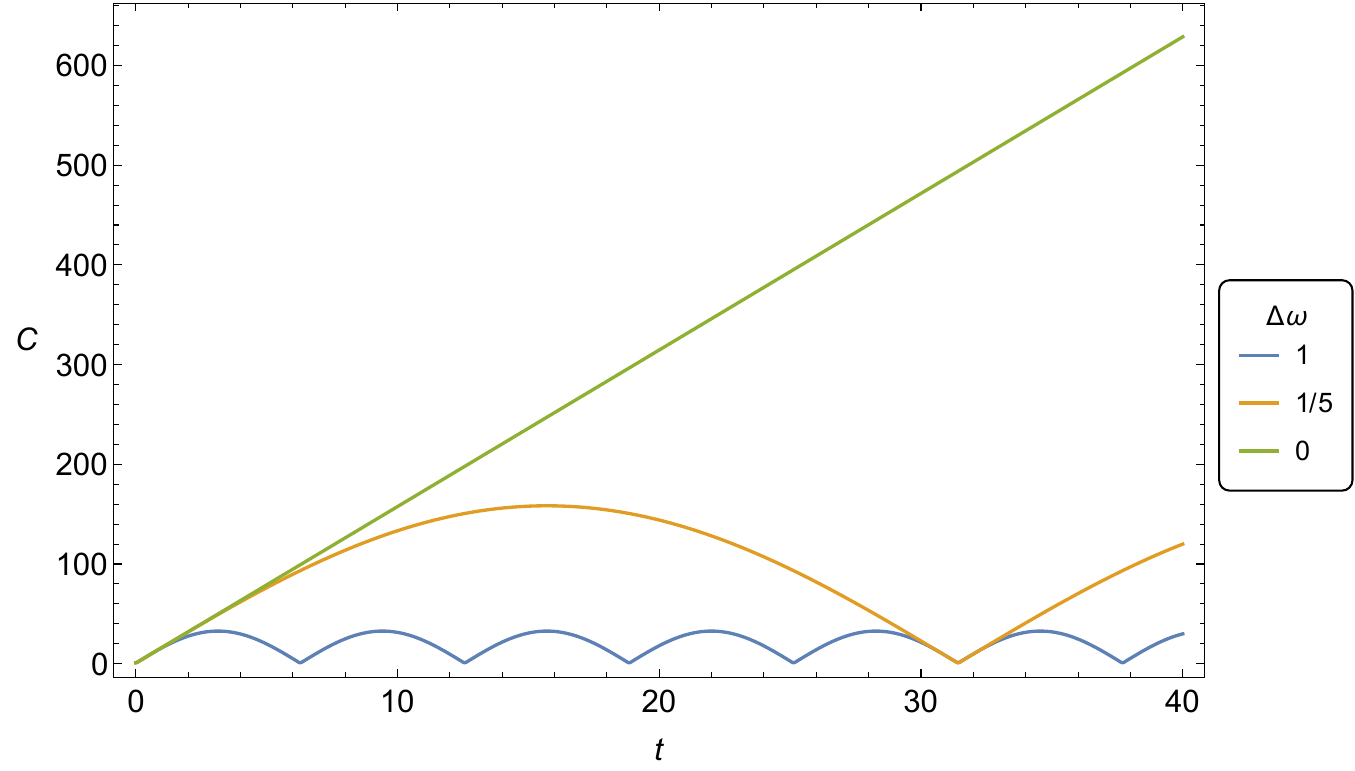}
        \subcaption{Krylov complexity for various $\Delta\omega$.}
        \label{fig:rectangular-Krylov}
      \end{minipage} \\
    \caption{The behaviors of Lanczos coefficients and Krylov complexity when varying $\Delta\omega$ with $\omega_{\rm max}=20$ and $\alpha=\infty$.}
\end{figure}

If we set $\alpha=\infty$ instead of $\alpha=1$, the power spectrum becomes a sum of uniform delta function peaks.
The Lanczos coefficients become as Fig.~\ref{fig:rectangular-Lanczos}, and the initial linear growth disappears.
As when $\alpha=1$, the plateau of the Lanczos coefficients extends as $\Delta\omega / \omega_{\rm max}$ is decreased, and in the continuous limit $\Delta\omega / \omega_{\rm max} \ll 1$, the plateau continues forever as in \cite{viswanath1994recursion}.
For Krylov complexity (Fig.~\ref{fig:rectangular-Krylov}), as in the case of $\alpha=1$, the range of linear increase becomes longer as $\Delta\omega / \omega_{\rm max}$ is decreased.

\subsubsection{Varying $\omega_{\rm max}$}
\begin{figure}[t]
      \begin{minipage}[t]{0.45\hsize}
        \centering
        \includegraphics[width=.95\columnwidth]{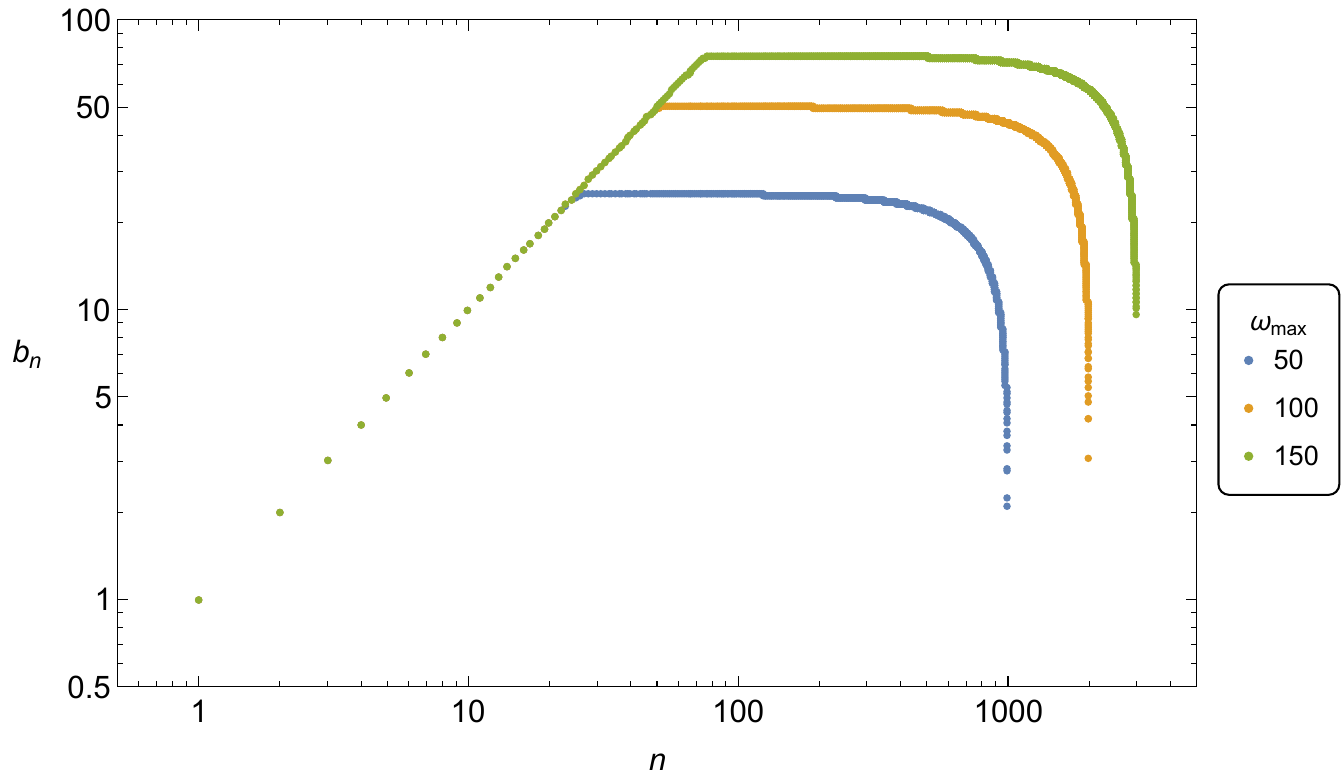}
        \subcaption{Lanczos coefficients for various $\omega_{\rm max}$.}
        \label{fig:omega-max-Lanczos}
      \end{minipage}
      \hspace{5mm}
      \begin{minipage}[t]{0.45\hsize}
        \centering
        \includegraphics[width=.95\columnwidth]{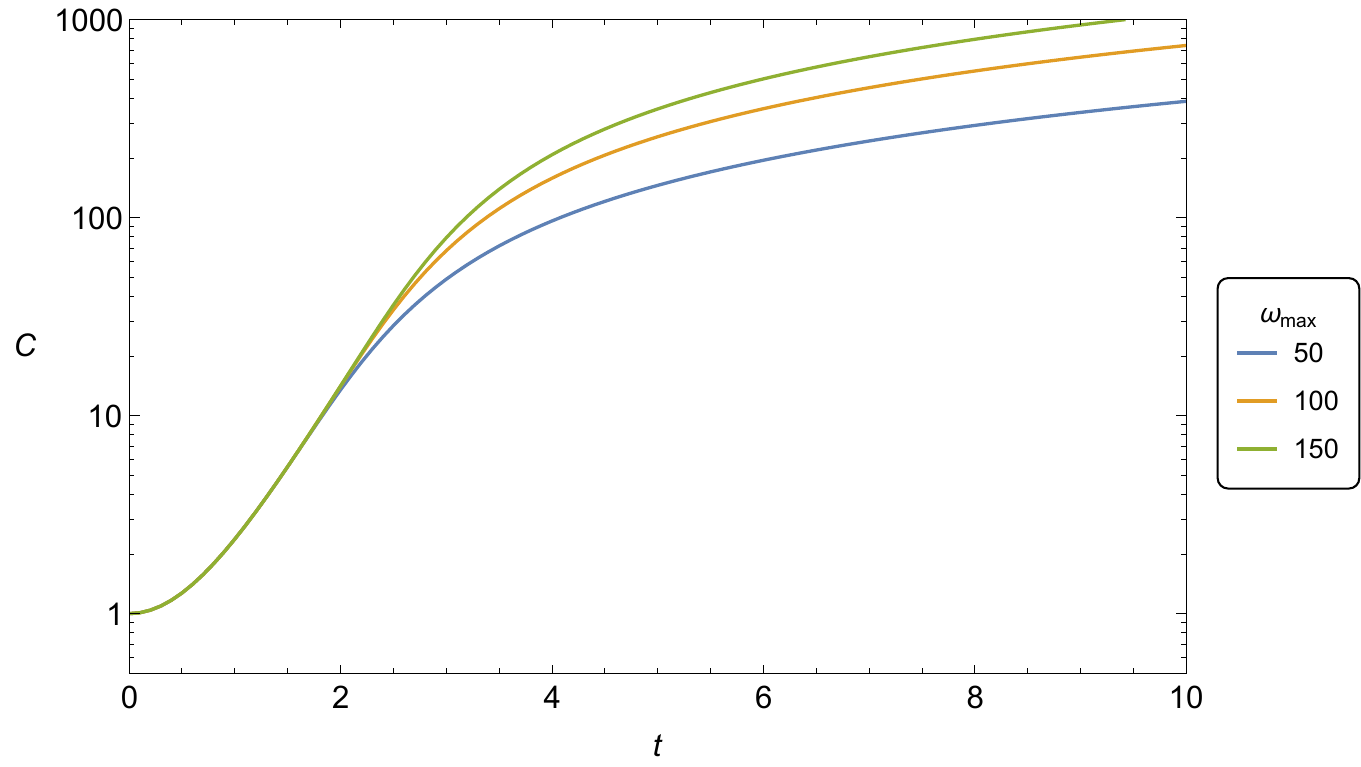}
        \subcaption{Krylov complexity for various $\omega_{\rm max}$.}
        \label{fig:omega-max-Krylov}
      \end{minipage} \\
    \caption{The behaviors of Lanczos coefficients and Krylov complexity when varying $\omega_{\rm max}$ with $\Delta\omega=1/10$ and $\alpha=1$.}
\end{figure}

When $\Phi(\omega)$ is continuous, it is known that the width of the support of $\Phi(\omega)$ determines the asymptotic value of the Lanczos coefficient.
Here we examine this for the discrete case by varying $\omega_{\rm max}$ in \eqref{eq:toy-power-spectrum} while $\Delta\omega=1/10$ and $\alpha=1$.
In Fig.~\ref{fig:omega-max-Lanczos}, we show the resulting Lanczos coefficients.
For each $\omega_{\rm max}$, the Lanczos coefficients seem to grow linearly in the beginning, then reach a constant value.
Note that, strictly speaking, the apparent linear growth in \ref{fig:omega-max-Lanczos} should include a small amount of staggering unless $\Delta\omega / \alpha$ is sufficiently small.
The constant value $b$, which is reached after linear growth, is determined as $b=\omega_{\rm max}/2$ by the size of the support $[-\omega_{\rm max},\omega_{\rm max}]$ as in the continuous case.
Since the number of delta function peaks is finite, the Lanczos coefficients eventually decay to zero.

Of particular interest is the fact that the range where the Lanczos coefficient grows linearly becomes longer as $\omega_{\rm max}$ becomes larger.
Correspondingly, in Fig.~\ref{fig:omega-max-Krylov}, Krylov complexity also appears to extend the range of exponential growth.

In the limit of $\omega_{\rm max}\to \infty$, the Lanczos coefficients only increase linearly, and Krylov complexity is expected to increase exponentially forever if $\Delta\omega / \alpha$ is sufficiently small.

\subsubsection{Varying $\alpha$}
\begin{figure}[t]
      \begin{minipage}[t]{0.45\hsize}
        \centering
        \includegraphics[width=.95\columnwidth]{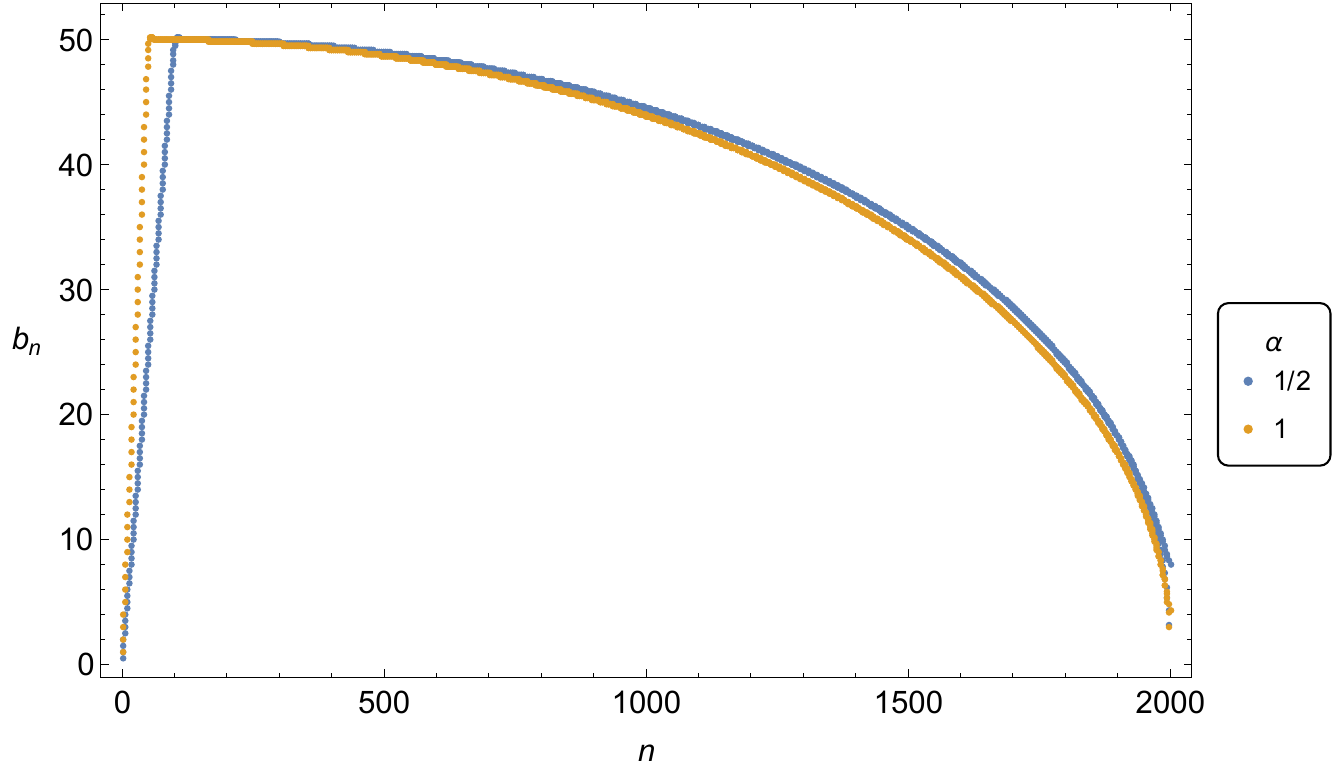}
        \subcaption{Lanczos coefficients for various $\alpha$.}
        \label{fig:whole-Lanczos}
      \end{minipage}
      \hspace{5mm}
      \begin{minipage}[t]{0.45\hsize}
        \centering
        \includegraphics[width=.95\columnwidth]{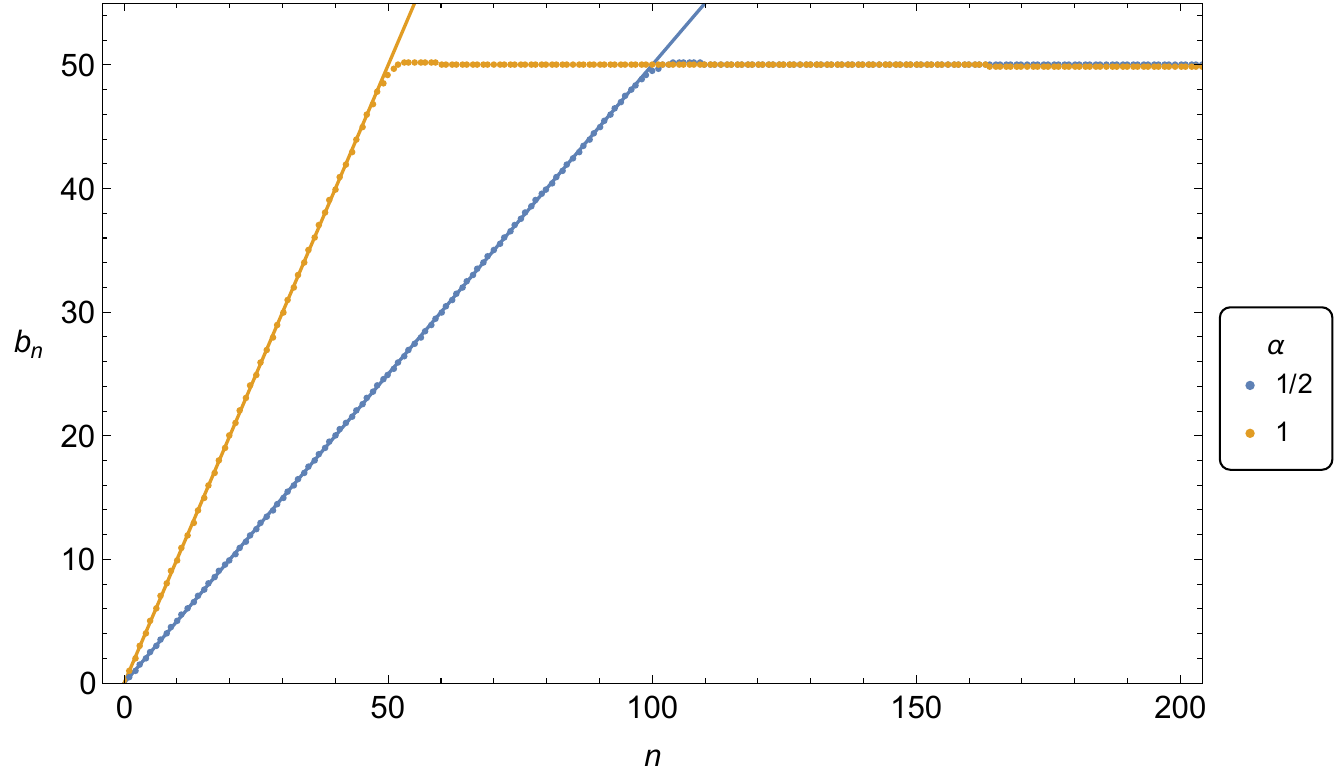}
        \subcaption{Enlarged initial portion of (a).}
        \label{fig:initial-Lanczos}
      \end{minipage} \\
    \caption{The behaviors of Lanczos coefficients when varying $\alpha$ with $\omega_{\rm max}=100$ and $\Delta \omega=1/10$.}
    \label{fig:Lanczos-alpha}
\end{figure}

\begin{figure}[t]
      \begin{minipage}[t]{0.45\hsize}
        \centering
        \includegraphics[width=.95\columnwidth]{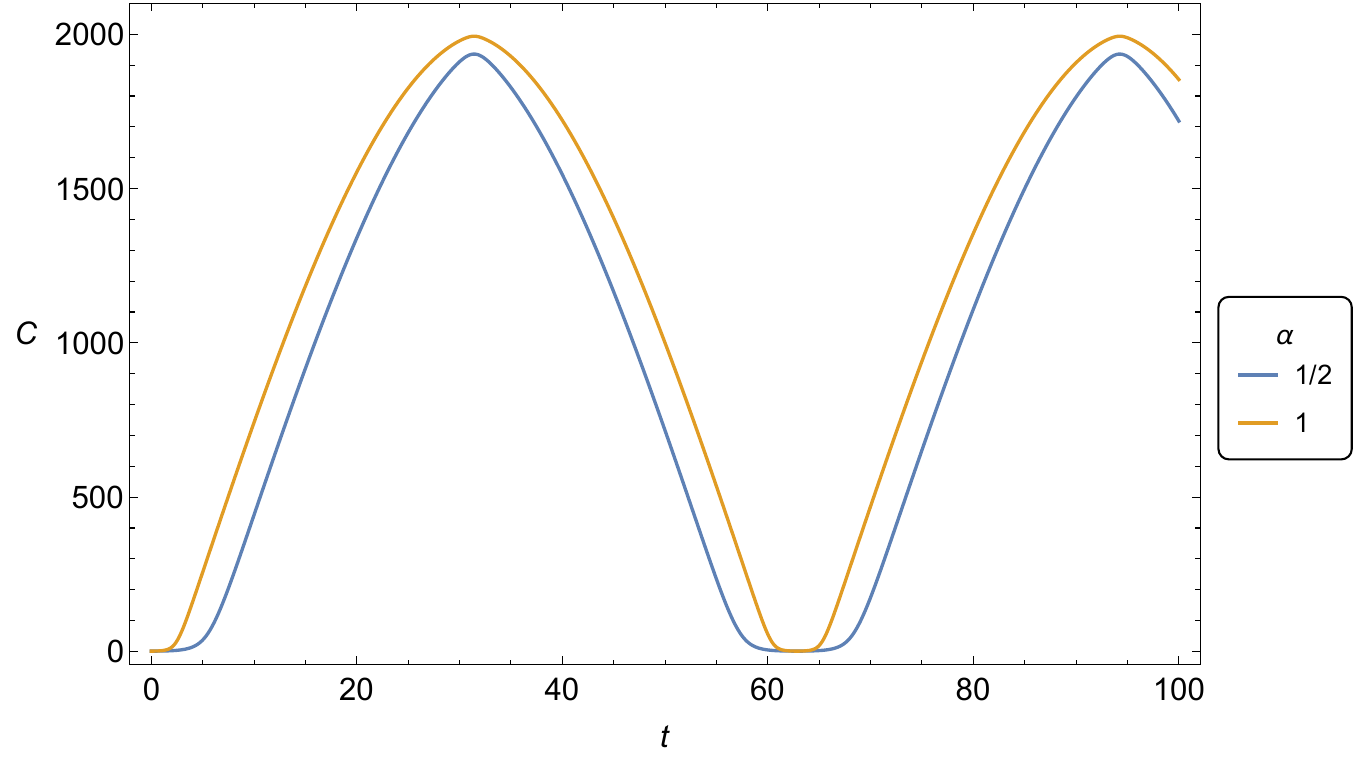}
        \subcaption{Krylov complexity for various $\alpha$.}
        \label{fig:whole-Krylov}
      \end{minipage}
      \hspace{5mm}
      \begin{minipage}[t]{0.45\hsize}
        \centering
        \includegraphics[width=.95\columnwidth]{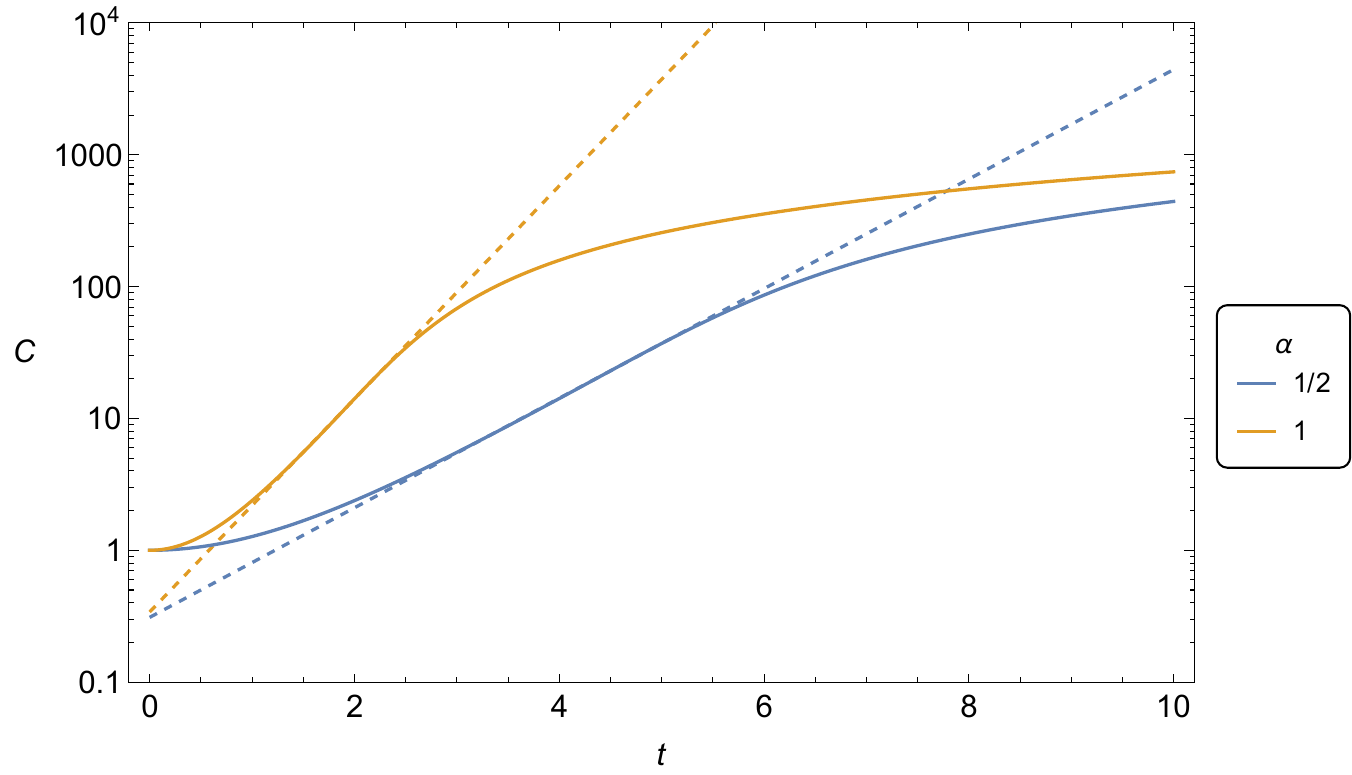}
        \subcaption{Enlarged initial portion of (a).}
        \label{fig:initial-Krylov}
      \end{minipage} \\
    \caption{The behaviors of Krylov complexity when varying $\alpha$ with $\omega_{\rm max}=100$ and $\Delta \omega=1/10$.}
    \label{fig:Krylov-alpha}
\end{figure}

Next, we fix $\omega_{\rm max}$ and $\Delta\omega$ and vary $\alpha$.
The resulting Lanczos coefficients are shown in Fig.~\ref{fig:Lanczos-alpha}.
When $\alpha$ is varied, the slope of the initial linear growth of the Lanczos coefficient changes. This is consistent with $\alpha$ acting as the temperature of this spectrum, as discussed earlier.
The solid lines in Fig.~\ref{fig:initial-Lanczos} are $b_n=\alpha n$.
This is in line with the statement around \eqref{eq:sech-power-spectrum}.
Correspondingly, the exponential growth rate of Krylov complexity also changes.
In Fig.~\ref{fig:Krylov-alpha} we show the time dependence of Krylov complexity.
The dashed lines in Fig.~\ref{fig:initial-Krylov} are the results of the fitting, each of which is found to be $0.310 \times \exp(0.957t)$ when $\alpha = 1/2$ and $0.340 \times \exp(1.86t)$ when $\alpha = 1$.
This is relatively consistent with the Krylov complexity behaving like $e^{2\alpha t}$ in late time when the Lanczos coefficient is perfectly linear $b_n = \alpha n$.

\subsection{Constraints from the energy spectrum}
\label{sec:spectrum-support}
As we have already used very extensively, the Lanczos coefficients and Krylov complexity can be computed from the power spectrum.
However, they depend on the choice of operator.
A system-specific concept that does not depend on the choice of operator is the density of states.
If we can impose general constraints on the power spectrum based on the behavior of the density of states, we can discuss the Lanczos coefficient and Krylov complexity from a more general viewpoint.
In this regard, we note the following general property.
\begin{description}
    \item[Claim] Let $H$ be the Hamiltonian, $|E_i\rangle$ be the energy eigenstate of energy $E_i$ and $\sigma(H)$ be the set of all energy eigenvalues of $H$. If $\sigma(H)$ is bounded, then the power spectrum of the auto-correlation function $C(t)={\rm tr}(\mathcal{O}(t)\mathcal{O}(0))$ of an arbitrary operator $\mathcal{O}$ must have a bounded support, and vice versa.
\end{description}
This property is obvious if we write down the auto-correlation function using the energy eigenbasis as has been done in \cite{Camargo:2022rnt}.
For systems with a bounded spectrum, the power spectrum of any operator has a bounded support and, in particular, no tail, so the Krylov complexity does not increase exponentially in late time.
Conversely, the exponential growth in Krylov complexity at late time is allowed only when the energy spectrum is unbounded.
For instance, in the DSSYK model with $\lambda>0$, the energy spectrum is bounded, so Krylov complexity of any operator does not grow exponentially in late time.
On the other hand, in the large-$p$ SYK model with $N\to\infty$, the energy spectrum is unbounded, so Krylov complexity of a single fermion operator is allowed to grow exponentially in late time \cite{Parker:2018yvk}.



\section{Discussion}
\label{sec:discussion}
In this paper, we have studied Krylov complexity of the fermion chain operators of the DSSYK model in various parameter regions and confirmed its exponential growth.
In particular, the increasing exponent saturates the chaos bound, confirming that the prediction that the exponential growth rate of Krylov complexity provides an upper bound on the exponential behavior of the OTOC is indeed true for the DSSYK model in particular regions.
Krylov complexity can be completely determined from the auto-correlation function of the operator and is fully characterized by the power spectrum.
In the case of continuous spectra, their correspondence was well known \cite{Parker:2018yvk,viswanath1994recursion,Barbon:2019wsy,Camargo:2022rnt}.
We studied the power spectrum of the fermion chain operator in the DSSYK model and gave an understanding of the behavior of the Lanczos coefficients and Krylov complexity.
In particular, regarding the condition on the structure of staggering often seen in Lanczos coefficients, we discussed the possibility of adding the finiteness of the higher-order derivative to the conditions on the derivative of the power spectrum proposed by \cite{Camargo:2022rnt}.
Moreover, considering the time when Krylov complexity changes from exponential growth to linear growth as the scrambling time, we also discussed that in the DSSYK model, the scrambling time does not depend on the number $N$ of degrees of freedom in the system.
In this sense, the DSSYK model is a {\it hyperfast scrambler}.

Furthermore, by using an toy power spectrum, we have obtained a systematic understanding of the behavior of the Lanczos coefficient.
Depending on whether the levels are discrete or continuous, the behavior of the Lanczos coefficients can differ in two ways.
The first is the difference in staggering of the Lanczos coefficients caused by the degree of the discreteness. 
Even when the energy spectrum is discrete, if the bulk of the power spectrum is sech-like, the Lanczos coefficient can have an initial linear growth. 
This slope is roughly the typical energy scale of the system (e.g., temperature), and if the discreteness of the levels is larger compared to this energy scale, staggering can occur in the Lanczos coefficients. 
On the other hand, if the discreteness is sufficiently small, the initial behavior of the Lanczos coefficients is almost indistinguishable from the continuous case.
The second difference is the asymptotic behavior of the Lanczos coefficient, which depends on whether the number of levels is finite or infinite.
In the finite system, the dimension of the Krylov subspace is also finite, and the Lanczos coefficient eventually becomes zero. However, note that even if the levels are discrete, the Lanczos coefficients can continue to increase if the number of levels is infinite.
Also, the support of the power spectrum determines the plateau value of the Lanczos coefficient.
In particular, if the energy spectrum is bounded, the power spectrum is also bounded, so the growth in the Lanczos coefficient always stops eventually.
Since the plateau of the Lanczos coefficient corresponds to a linear increase in Krylov complexity, the energy spectrum must be bounded in order for Krylov complexity to grow linearly at a late time.

Let us comment on the relationship between the chaotic nature of a given system and the Lanczos coefficients and Krylov complexity.
In \cite{Parker:2018yvk}, it was shown for quantum many-body systems that Lanczos coefficient does not increase faster than linear increase.
It was also conjectured that asymptotic linear growth of the Lanczos coefficients is related to quantum chaos.
These arguments are for quantum many-body systems and focus on the tail behavior of the power spectrum.
If we consider a quantum system with finite degrees of freedom, the power spectrum has no tail, and the Lanczos coefficients decay asymptotically to zero because the dimension of Krylov subspace is finite. 
The initial linear growth of the Lanczos coefficients is, as we have seen with the toy power spectrum, a result of the sech-like behavior of the bulk portion of the power spectrum.
However, the detailed shape of the bulk of the power spectrum and the initial growth regime of the Lanczos coefficients are highly dependent on the choice of operator.
On the other hand, the behavior of the Lanczos coefficients after the initial growth regime did not change significantly when the shape of the toy power spectrum and the choice of operator were changed. 
This suggests that looking at the Lanczos coefficients after the initial growth can provide system-specific properties.
Traditionally, the statistical distribution of level spacing has been used to characterize quantum chaotic properties \cite{Bohigas:1983er}.
This characterization can be applied not only to quantum many-body systems but also to finite-dimensional quantum systems.
In a real system, the energy spectrum is not equally spaced but fluctuates, and the power spectrum becomes the sum of delta function peaks distributed at various intervals.
This fluctuation is expected to affect the late time behavior of the Lanczos coefficients \cite{Rabinovici:2021qqt, Rabinovici:2022beu, Hashimoto:2023swv, Bhattacharyya:2023dhp, Menzler:2024atb, Scialchi:2023bmw}.

In this paper, we have considered operator complexity.
On the other hand, the complexity of quantum states has long been of interest because it is expected to correspond to wormhole volumes and the like in the AdS/CFT correspondence \cite{Susskind:2014rva,Stanford:2014jda,Brown:2015bva,Brown:2015lvg}.
Recently, Krylov complexity for quantum states has also been proposed and studied \cite{Balasubramanian:2022tpr}.
This complexity is a natural extension of the operator case definition.\footnote{There has also been a recent proposal to assemble quantum states into a density matrix and consider the Krylov complexity of the density matrix \cite{Caputa:2024vrn}.}
However, it is not clear whether it is possible to characterize it using quantities corresponding to the power spectrum in the operator case.
The definition of complexity on the quantum theory side, which is a dual to the holographic complexity on the bulk gravity theory side that has been studied in the past, is still unclear.\footnote{In \cite{Rabinovici:2023yex}, Krylov complexity of the chord state of the DSSYK model in the high-temperature limit was studied and its time dependence was found to be consistent with the time dependence of the volume (geodesic length) of the wormhole connecting the two asymptotic regions of a two-dimensional black hole. It remains to be confirmed whether this is also true for more general setups.}
It is an important issue to obtain a systematic understanding of the Krylov complexity and Lanczos coefficients of quantum states as in the case of operators.

It is also future work to give a bulk-side interpretation to the Krylov complexity of the operator itself.\footnote{In another direction, studies have been conducted to interpret Krylov complexity geometrically using information metrics \cite{Caputa:2021sib}.}
It has long been proposed that the size of the operator in a quantum system at the boundary corresponds to the momentum of the bulk particle \cite{Susskind:2018tei}.
It would be interesting to consider whether we can embed the Krylov complexity in this conjecture.

\section*{Acknowledgments}
We would like to thank Mitsuhiro Nishida and Norihiro Tanahashi for helpful comments on our draft. The work of R.~W.~was supported by Grant-in-Aid for JSPS Fellows No.~JP22KJ1940.

\appendix
\section{Justification for Poisson approximation}
\label{app:JPA}
The probability distribution of $k$ fermions being the same when the chord is crossed is
\begin{align}
P(k) &= \frac{1}{\begin{pmatrix}N \\ p\end{pmatrix}} \begin{pmatrix}p \\ k\end{pmatrix} \cdot \begin{pmatrix}N-p \\ p-k\end{pmatrix}
\end{align}
If this can be Poisson approximated under $k\ll p \ll N$, then the argument in \cite{Berkooz:2018jqr} follows. Here, we examine more rigorously the parameter regions for which the approximation can be justified, giving specifically what hierarchy is desired, for example, $p \sim O(N^{\#})$.\\

\subsection{Poisson approximation}
Consider the situation where $k\ll p \ll N$. Expanding with large limit $p$ as $k=O(p^\alpha)\ (\frac{k}{p^\alpha} \equiv \lambda_2)$, we obtain
\begin{align}
\frac{p!}{(p-k)!} &= p^k e^{-\frac{1}{2}\lambda_2^2 \frac{1}{p^{1-2\alpha}}} \left( 1 + \frac{1}{2} \lambda_2 \frac{1}{p^{1-\alpha}} + \cdots \right)
 \end{align}
Therefore, if $1-2\alpha>0\ \to\ \alpha<1/2$, in other words, $k\sim o(\sqrt{p})$, we can approximate
\begin{align}
P(k) \sim \frac{\begin{pmatrix}N-p \\ p\end{pmatrix}}{\begin{pmatrix}N \\ p\end{pmatrix}} \frac{p^{2k}}{k!} \frac{(N-2p)!}{(N-2p+k)!}
\end{align}
In the same reason, if $k\sim o(\sqrt{N-2p})$, we can approximate
\begin{align}
P(k) \sim \frac{\begin{pmatrix}N-p \\ p\end{pmatrix}}{\begin{pmatrix}N \\ p\end{pmatrix}} \frac{p^{2k}}{k!} \frac{1}{(N-2p)^k}
\end{align}
Let us take the logarithm,
\begin{align}
\log P(k) = \log \frac{p^{2k}}{ k!} -k\log (N-2p) + \log \frac{((N-p)!)^2}{N!(N-2p)! }
\end{align}
If we set $p = o(N)$, the second term in rhs can be approximate
\begin{align}
k\log N + k\log\left( 1 - \frac{p}{N} \right) \simeq k\log N - k \frac{p}{N} +\cdots
\end{align}
For the remaining terms, we can use Stirling's formula $\log n! \sim n\log n - n$. Since Stirling's formula can be used for $n \gg 1$, if $p=o(N)$, then
\begin{align}
\log& \frac{((N-p)!)^2}{N!(N-2p)! } \sim 2(N-p)\log (N-p) -N\log N -(N-2p)\log(N-2p)\notag \\
&=  2\log(N-p)\log \left(1- \frac{p}{N}\right) -(N-2p)\log\left( 1 - \frac{2p}{N}\right) \sim N\left( -\frac{p^2}{N^2}-\frac{p^3}{N^3} +\cdots \right)
\end{align}
Therefore, the probability distribution can be approximated by
\begin{align}
P(k) \sim \frac{p^{2k}}{ N^k k!} e^{-N\left( \frac{p^2}{N^2}+\frac{p^3}{N^3} +\cdots \right)} ,\ \ p\sim o(N),\ \ k\sim o(\sqrt{p})
\end{align}

\subsection{Evaluating the peak point}
Since $k$ can essentially take values from $1$ to $p$, the region of $k$ in which this Poisson approximation can be justified is not large. However, the original probability distribution and the approximated Poisson distribution have the characteristic that there exists a peak point, otherwise it approaches zero rapidly. Therefore, the peak value is the dominant contribution to the average.\\

From the above, another important issue to be discussed is whether the peak of the original probability distribution exists within the range where the Poisson approximation can be justified.\\

The peak of this distribution is obtained by evaluating the following
\begin{align}
\frac{P(k)}{P(k+1)} = \frac{N(k+1)}{p^2} 
\end{align}
Thus, it decreases if $k>\frac{p^2}{N}-1$. In other words, the maximum value is obtained near this point. This peak must be well contained within $k \sim o(\sqrt{p})$, which can be approximated by Poisson. From the above, the following hierarchy
\begin{align}
\frac{p^2}{N} \ll \sqrt{p}
\end{align}
must exist. In the case of the conventional double-scaled SYK model, $p\sim O(\sqrt{N})$. The Poisson distribution can be sufficiently approximated in the range of $k \sim o(N^{1/4})$. In addition, at this time, since
\begin{align}
\frac{p^2}{N}\ (= O(1)) \ll \sqrt{p}\ ( = O(N^{1/4}))
\end{align}
we can justify Poisson approximation.\\

In general, when we consider a particular scaling limit $p \sim O(N^x)$, the Poisson distribution can be sufficiently approximated within $k \sim o(N^{x/2})$. At this time, the Poisson approximation is justified if there is a hierarchy of
\begin{align}
\frac{p^2}{N}\ (= O(N^{2x-1})) \ll \sqrt{p}\ ( = O(N^{x/2}))
\end{align}
Therefore, this approximation is valid in the range $2x-1<x/2\ \to\ x<2/3$.\\

From the above, if we take the limit $\frac{2 p^\alpha}{N} \equiv \lambda$ fixed, the method in \cite{Berkooz:2018jqr} is justified if $\alpha > \frac{3}{2}$. At this time
\begin{align}
P(k) \sim \frac{1}{ k!} \left(\frac{1}{2}\lambda \right)^{\frac{2k}{\alpha}} N^{(\frac{2}{\alpha}-1)k} e^{-\left(\frac{1}{2}\lambda \right)^{\frac{2}{\alpha}}N^{\frac{2}{\alpha}-1}\left( 1 + \left(\frac{1}{2}\lambda \right)^{\frac{1}{\alpha}}N^{\frac{1}{\alpha}-1} +\cdots \right) }
\end{align}
Thus, the expectation value of $(-1)^k$ becomes
\begin{align}
q 
&\equiv \sum_k (-1)^k \frac{1}{ k!} \left(\frac{1}{2}\lambda \right)^{\frac{2k}{\alpha}} N^{(\frac{2}{\alpha}-1)k} e^{-\left(\frac{1}{2}\lambda \right)^{\frac{2}{\alpha}}N^{\frac{2}{\alpha}-1}\left( 1 + \left(\frac{1}{2}\lambda \right)^{\frac{1}{\alpha}}N^{\frac{1}{\alpha}-1} +\cdots \right) } \notag \\
&= e^{-\left(\frac{1}{2}\lambda \right)^{\frac{2}{\alpha}}N^{\frac{2}{\alpha}-1}\left( 2 + \left(\frac{1}{2}\lambda \right)^{\frac{1}{\alpha}}N^{\frac{1}{\alpha}-1} +\cdots \right) }
\label{eq:q}
\end{align}
When $\alpha = 2$, $q = e^{-\lambda}$ fixed. When $\frac{1}{2}<\alpha<\frac{2}{3}$, $q \sim e^{-N^\#}\to 0\,, (\#>0)$.


\bibliographystyle{JHEP}
\bibliography{reference.bib}
\end{document}